\documentclass[twocolumn]{aastex631}
\usepackage{amsmath}
\usepackage{multirow}

\shorttitle{WASP-76b with IGRINS}
\shortauthors{Weiner Mansfield et al.}

\begin{document}

\title{The metallicity and carbon-to-oxygen ratio of the ultra-hot Jupiter WASP-76b from Gemini-S/IGRINS}

\correspondingauthor{Megan Weiner Mansfield}
\email{meganmansfield@arizona.edu}

\author[0000-0003-4241-7413]{Megan Weiner Mansfield}
\affiliation{Steward Observatory, University of Arizona,
Tucson, AZ 85719, USA}
\affiliation{NHFP Sagan Fellow}

\author{Michael R. Line}
\affiliation{School of Earth and Space Exploration, Arizona State University, Tempe, AZ 85281, USA}

\author{Joost P. Wardenier}
\affiliation{Department of Physics (Atmospheric, Oceanic, and Planetary Physics), University of Oxford, Oxford, OX1 3PU, UK}

\author{Matteo Brogi}
\affiliation{Dipartimento di Fisica, Universit\'a degli Studi di Torino, via Pietro Giuria 1, I-10125, Torino, Italy}

\author[0000-0003-4733-6532]{Jacob L. Bean}
\affiliation{Department of Astronomy \& Astrophysics, University of Chicago, Chicago, IL 60637, USA}

\author{Hayley Beltz}
\affiliation{Department of Astronomy, University of Maryland, College Park, MD 20742, USA}

\author{Peter Smith}
\affiliation{School of Earth and Space Exploration, Arizona State University, Tempe, AZ 85281, USA}

\author{Joseph A. Zalesky}
\affiliation{Department of Astronomy, The University of Texas at Austin, Austin, TX 78712, USA}

\author[0000-0003-1240-6844]{Natasha Batalha}
\affiliation{NASA Ames Research Center, Moffett Field, CA 94035, USA}

\author{Eliza M.-R. Kempton}
\affiliation{Department of Astronomy, University of Maryland, College Park, MD 20742, USA}

\author{Benjamin T. Montet}
\affiliation{School of Physics, University of New South Wales, Kensington, New South Wales, Australia}

\author{James E. Owen}
\affiliation{Astrophysics Group, Department of Physics, Imperial College London, London SW7 2AZ, UK}

\author{Peter Plavchan}
\affiliation{Department of Physics \& Astronomy, George Mason University, Fairfax, VA 22030}

\author{Emily Rauscher}
\affiliation{Department of Astronomy, University of Michigan, Ann Arbor, MI 48109, USA}

\begin{abstract}

Measurements of the carbon-to-oxygen (C/O) ratios of exoplanet atmospheres can reveal details about their formation and evolution. Recently, high-resolution cross-correlation analysis has emerged as a method of precisely constraining the C/O ratios of hot Jupiter atmospheres. We present two transits of the ultra-hot Jupiter WASP-76b observed between $1.4-2.4$~$\mu$m with the high-resolution Immersion GRating INfrared Spectrometer (IGRINS) on the Gemini-S telescope. We detected the presence of H$_{2}$O, CO, and OH at signal-to-noise ratios of $6.93$, $6.47$, and $3.90$, respectively. We performed two retrievals on this data set. A free retrieval for abundances of these three species retrieved a volatile metallicity of $\left[\frac{\mathrm{C}+\mathrm{O}} {\mathrm{H}}\right]=-0.70^{+1.27}_{-0.93}$, consistent with the stellar value, and a super-solar carbon-to-oxygen ratio of C/O$=0.80^{+0.07}_{-0.11}$. We also ran a chemically self-consistent grid retrieval, which agreed with the free retrieval within $1\sigma$ but favored a slightly more sub-stellar metallicity and solar C/O ratio ($\left[\frac{\mathrm{C}+\mathrm{O}} {\mathrm{H}}\right]=-0.74^{+0.23}_{-0.17}$ and C/O$=0.59^{+0.13}_{-0.14}$). 
A variety of formation pathways may explain the composition of WASP-76b. 
Additionally, we found systemic ($V_{sys}$) and Keplerian ($K_{p}$) velocity offsets which were broadly consistent with expectations from 3D general circulation models of WASP-76b, with the exception of a redshifted $V_{sys}$ for H$_{2}$O. Future observations to measure the phase-dependent velocity offsets and limb differences at high resolution on WASP-76b will be necessary to understand the H$_{2}$O velocity shift. Finally, we find that the population of exoplanets with precisely constrained C/O ratios generally trends toward super-solar C/O ratios. More results from high-resolution observations or \textit{JWST} will serve to further elucidate any population-level trends.

\end{abstract}

\keywords{}

\section{Introduction}
\label{sec:intro}

One of the main goals of transmission spectroscopy has been to use measurements of atmospheric compositions to understand the formation of hot Jupiters. For example, 
element ratios such as the carbon-to-oxygen (C/O) and silicon-to-oxygen (Si/O) ratio can provide information on their formation and evolution \citep[e.g.,][]{Oberg2011,Mordasini2016,Schneider2021,Schneider2021b,Molliere2022,Chachan2023}. More generally, the ratio of refractory to volatile elements can reveal the relative amounts of rocky and icy bodies accreted during formation \citep{Lothringer2021}.

Two decades of transmission spectroscopy on hot Jupiters have led to a wealth of information on their atmospheric compositions. The ultra-hot Jupiter WASP-76b in particular has been studied extensively through transits, eclipses, and full phase curves with the \textit{Hubble Space Telescope} (\textit{HST}) and \textit{Spitzer Space Telescope} \citep{vonEssen2020,Edwards2020,Fu2021,Mansfield2021,May2021}. More recently, the technique of high-resolution cross-correlation spectroscopy (HRCCS) has been used to detect a vast array of metals and volatile species in the atmosphere of WASP-76b \citep{Seidel2019,Ehrenreich2020,Casasayas2021,Deibert2021,Kesseli2021,Landman2021,Tabernero2021,Wardenier2021,Azevedo2022,Gandhi2022,Kawauchi2022,Kesseli2022,SanchezLopez2022,Savel2022,Deibert2023,Gandhi2023,Wardenier2023,Yan2023}. Ultimately, observations with \textit{HST}, \textit{Spitzer}, and VLT/CRIRES+ have placed some constraints on the abundances of the volatile species H$_{2}$O and CO. However, for all of these observations, the C/O ratio has been essentially unconstrained because none of them simultaneously detected all relevant carbon- and oxygen-bearing species at high significance.

In this paper we present a constraint on the C/O ratio of WASP-76b's atmosphere through observations of two transits of WASP-76b with the Immersion GRating INfrared Spectrometer (IGRINS) on the Gemini-S telescope. IGRINS is well-suited to these observations because of the combination of its high resolution ($R\approx45,000$) and large wavelength coverage ($1.4-2.4$~$\mu$m), which means it is sensitive to most of the primary oxygen- and carbon-bearing species in hot Jupiter atmospheres, such as H$_{2}$O, CO, and OH \citep[e.g.,][]{Line2021,Brogi2023}. In Section~\ref{sec:obs}, we describe the observations and data reduction. In Section~\ref{sec:ccor}, we use cross-correlations to detect the presence of H$_{2}$O, CO, and OH in the atmosphere of WASP-76b. In Sections~\ref{sec:retrieve} and \ref{sec:velret}, we use retrievals to constrain the abundances and the velocity offsets of the detected gases, respectively. In Section~\ref{sec:discuss}, we discuss these observations in the context of previous detections of carbon- and oxygen-bearing species in the atmosphere of WASP-76b and potential formation scenarios for this planet. Finally, we recap the main results and discuss future work in Section~\ref{sec:conclude}.

\section{Observations and Data Reduction}
\label{sec:obs}

We observed two transits of WASP-76b (mass $M_{p}=0.92M_{J}$, radius $R_{p}=1.83R_{J}$, period $P=1.81$~d, equilibrium temperature $T_{eq}=2160$~K) with Gemini-S/IGRINS on October 29, 2021 (night 1) and October 26, 2022 (night 2) as part of program GS-LP-107 (PI Mansfield). On the first and second nights we observed a sequence of 104 and 100 A-B pairs of exposures, respectively, with 77 pairs in transit on night 1 and 76 pairs in transit on night 2. Both observations used an exposure time of 45 seconds per exposure, or 90 seconds per A-B pair. The observations spanned orbital phases of $-0.057<\phi<0.056$ and $-0.059<\phi<0.054$, respectively, with phase 0 corresponding to mid-transit.

We used the IGRINS Pipeline Package \citep[PLP,][]{Sim2014,Lee2016} to reduce and optimally extract the spectra and perform an initial wavelength calibration. In order to separate out the signature of the transiting planet from the host star and telluric contamination, we next applied a custom pipeline, \texttt{IGRINS\_transit}\footnote{https://github.com/meganmansfield/IGRINS\_transit} \citep{igrins_transit}, based on the methods of \citet{Line2021}. This pipeline removes low signal-to-noise orders, performs a secondary wavelength calibration, and uses a singular value decomposition (SVD) to separate the planetary signal from the stellar and telluric signals.

We discarded orders 0, $20-27$, and $52-53$ from night 1 and orders 0, $22-26$, and $52-53$ from night 2 due to low transmittance and high telluric contamination due to their location at the edges of the H and K bands. We then performed a second wavelength calibration to correct for sub-pixel shifts in the wavelength solution over the course of an observation by applying a linear stretch and shift transform to each spectrum, using the spectrum observed closest to the on-sky wavelength calibration exposure as the template.

Following previous high-resolution studies \citep{deKok2013,Giacobbe2021,Line2021,Pelletier2021}, we used SVD to remove non-planetary contaminants. This method effectively identifies spectral features that are located on the same pixel over time, such as stellar absorption lines and telluric absorption, within the first few singular vectors (SVs), which can then be removed from the data to leave behind the planetary signal, which shifts across pixels over the course of a transit due to the redshift from the changing line-of-sight velocity as the planet orbits its host star. We used Python's \texttt{numpy.linalg.svd} function to perform the SVD and removed the first 4 SVs from the data to isolate the planetary signal. However, we found that repeating our analysis with removing 3 or 5 SVs did not change any of our results by more than $1\sigma$.

\section{Cross-Correlation Detections}
\label{sec:ccor}

Before doing a more in-depth retrieval analysis, we cross-correlated the observations with a range of models to identify which gases were detectable in our data. We used the model atmosphere framework described in \citet{Line2021} to create a solar composition, thermochemical equilibrium model at the equilibrium temperature of WASP-76b. To individually search for different gases, we artificially changed the gas abundances to only include H$_{2}$, He, and the single gas being searched for at a volume mixing ratio $10^{-3}$ (strong enough to present in the spectrum). We used the H$_{2}$O line list from POKAZATEL \citep{Polyansky2018,Gharib2021} and line lists for CO and OH from HITEMP \citep{Li2015,Gordon2022}. We combined the two nights of data by performing the cross-correlation for each night individually and summing the cross-correlation strengths. We then converted the cross-correlation strengths into detection signal-to-noise following 
\begin{equation}
\label{eq:sigma}
    SNR=(CCF-med)/stdev,
\end{equation}
where $SNR$ is the signal-to-noise, $CCF$ is the cross-correlation strength, and $med$ and $stdev$ are the $3\sigma$-clipped median and standard deviation calculated using \texttt{astropy.stats.sigma\_clipped\_stats}. Applying sigma clipping results in a more accurate SNR because it ignores the extended region of high cross-correlation strength surrounding the peak signal.

\begin{figure*}
    \centering
    \includegraphics[width=0.24\linewidth]{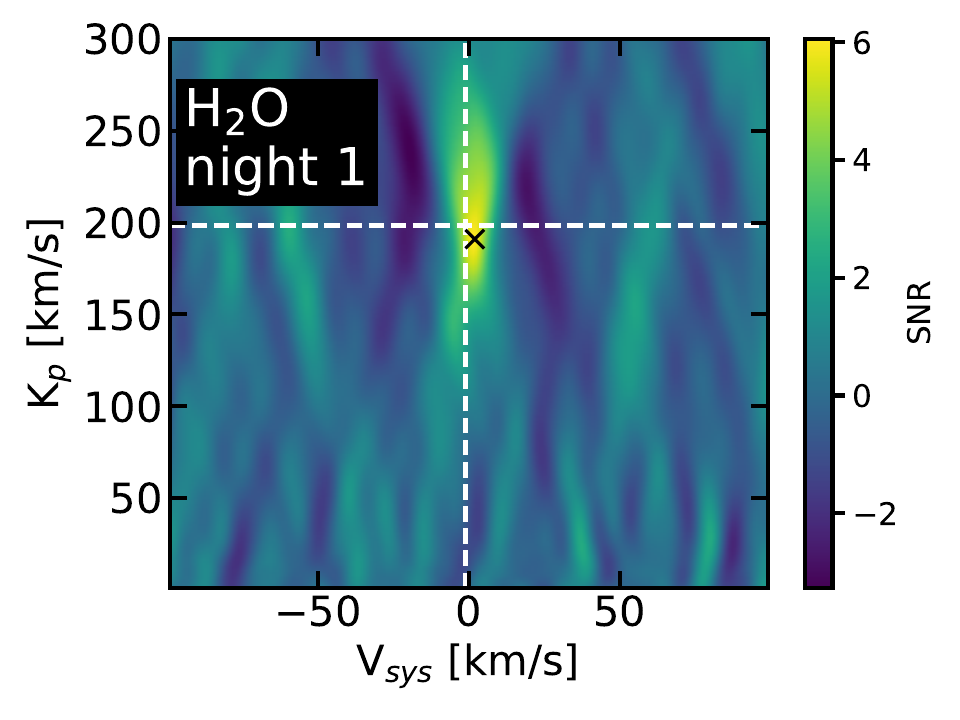}
    \includegraphics[width=0.24\linewidth]{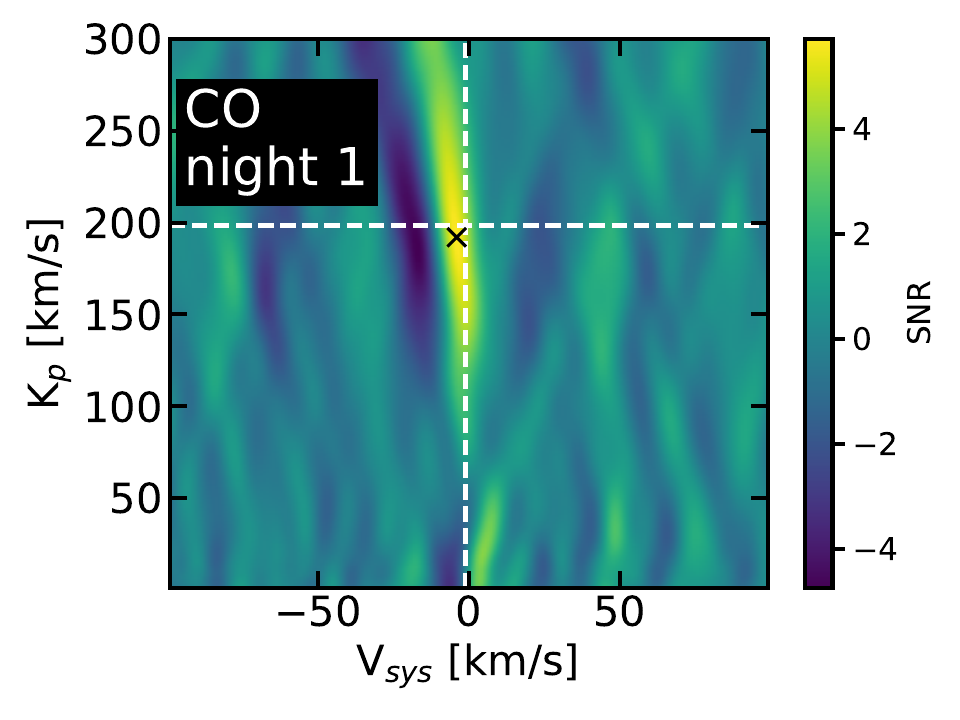}
    \includegraphics[width=0.24\linewidth]{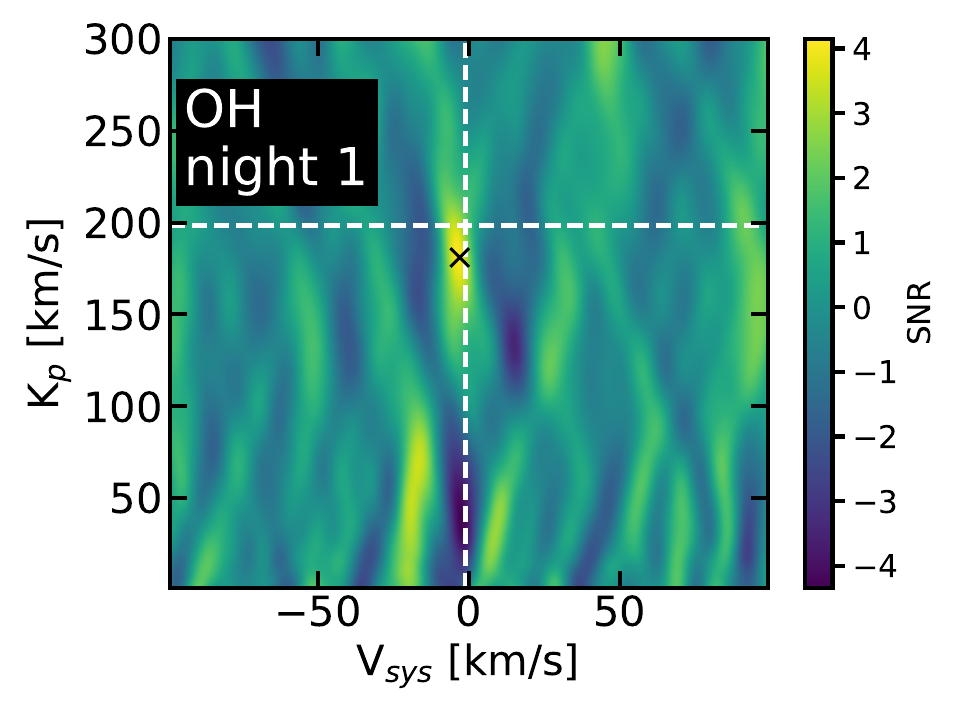}
    \includegraphics[width=0.24\linewidth]{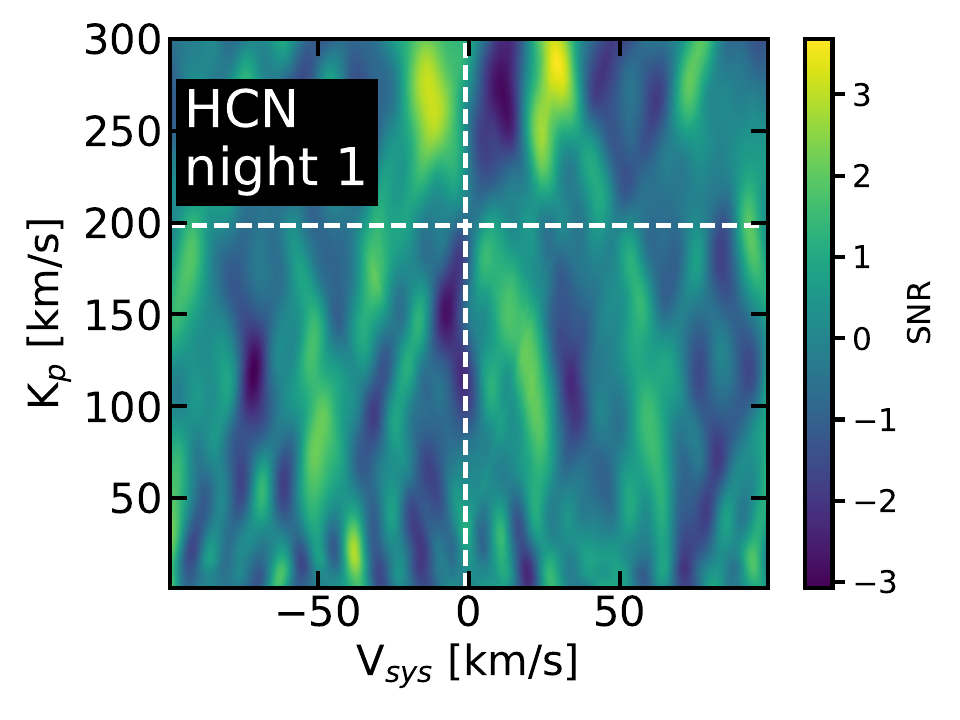}
    \includegraphics[width=0.24\linewidth]{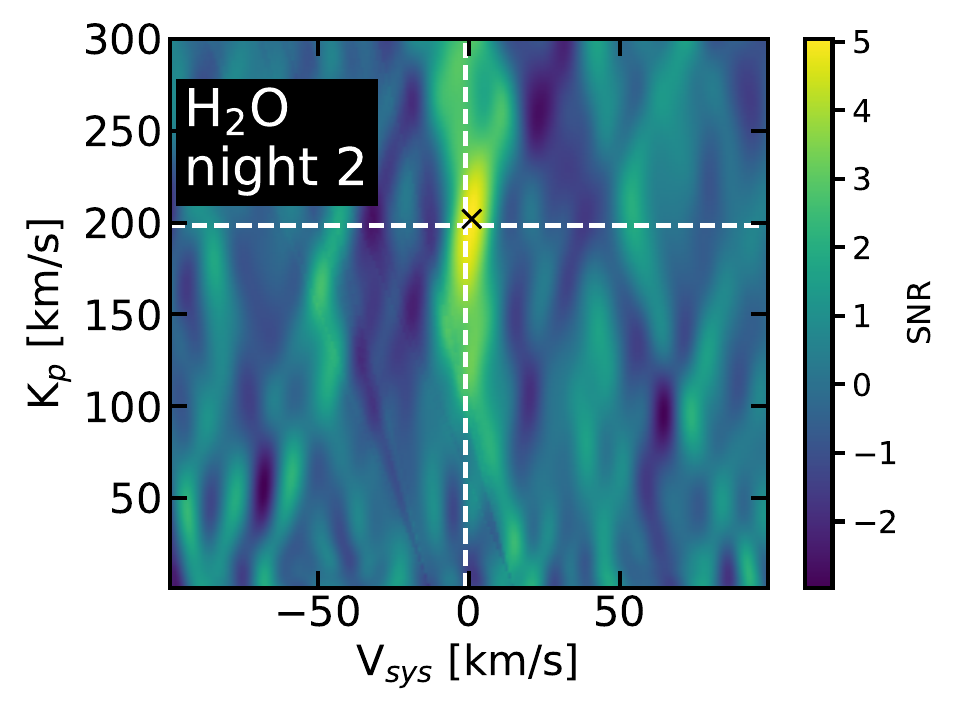}
    \includegraphics[width=0.24\linewidth]{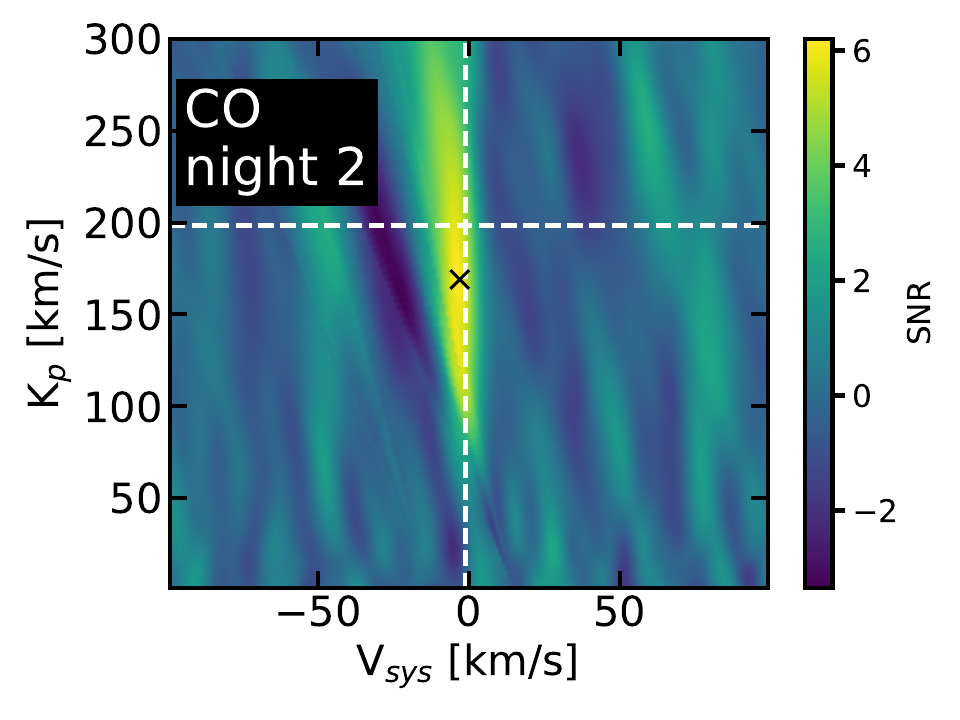}
    \includegraphics[width=0.24\linewidth]{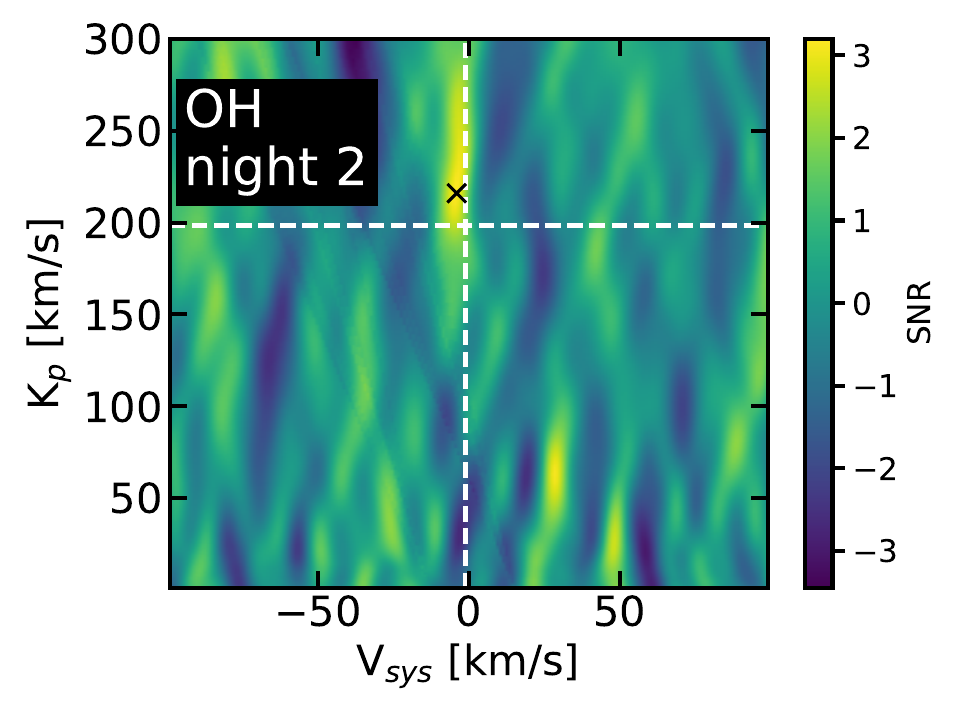}
    \includegraphics[width=0.24\linewidth]{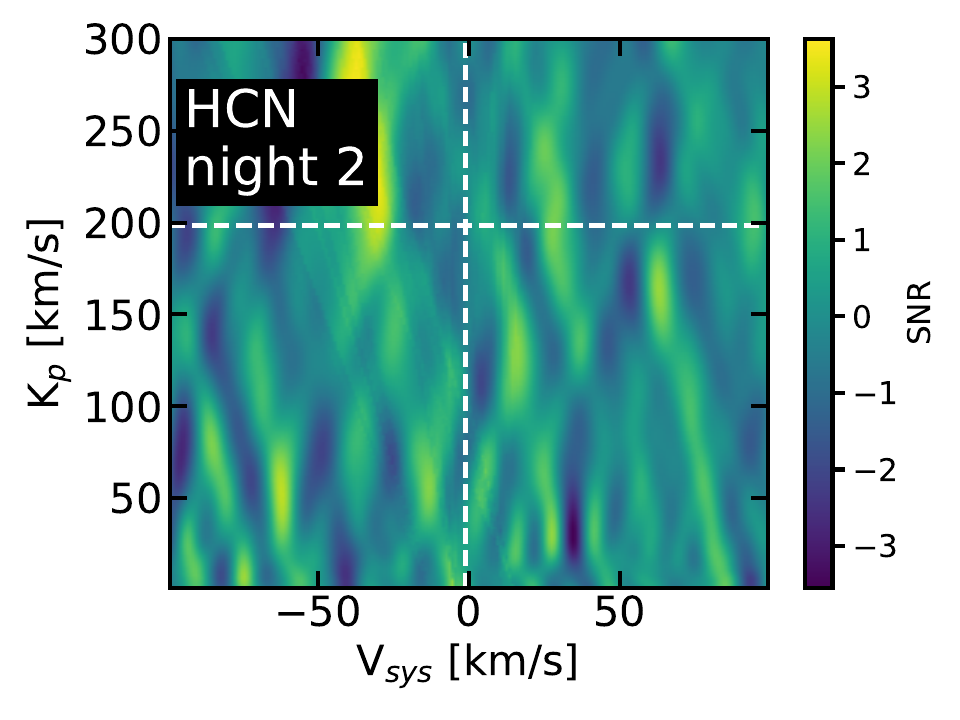}
    \includegraphics[width=0.24\linewidth]{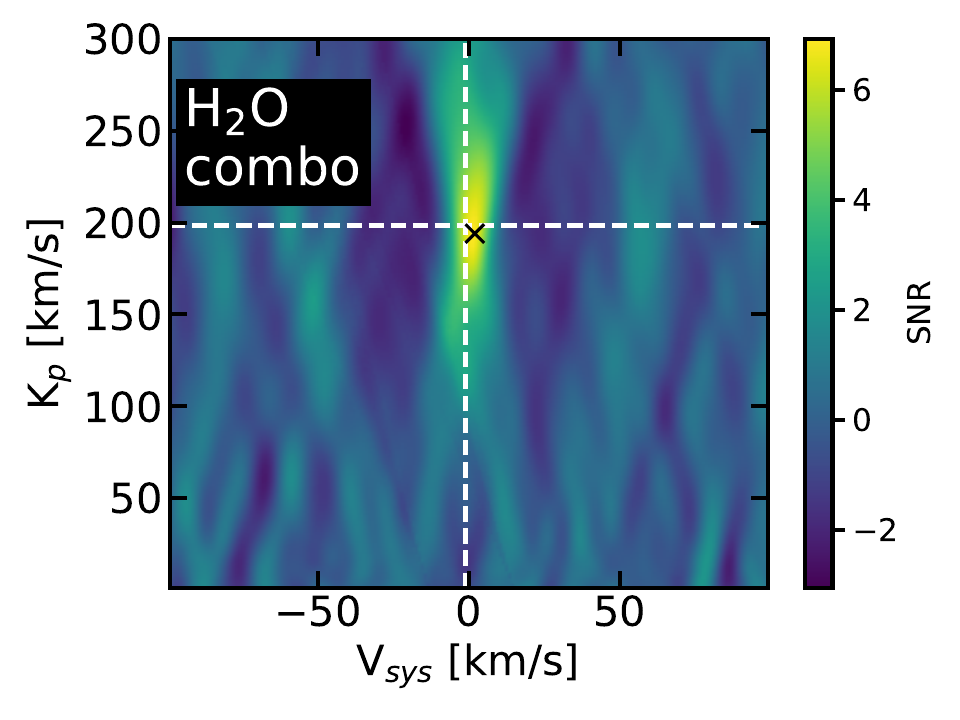}
    \includegraphics[width=0.24\linewidth]{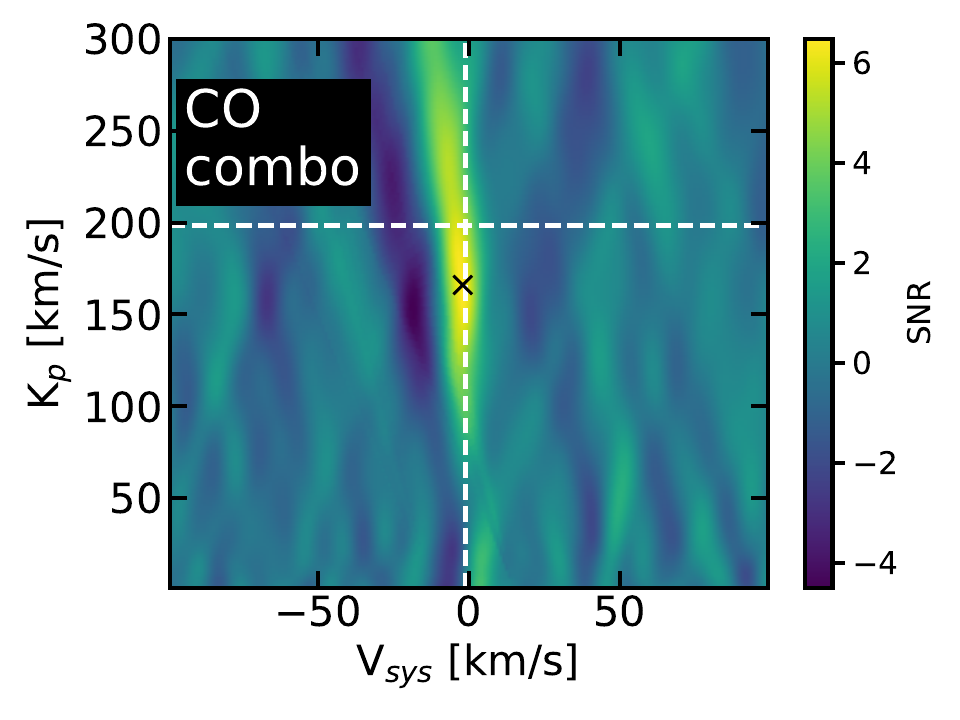}
    \includegraphics[width=0.24\linewidth]{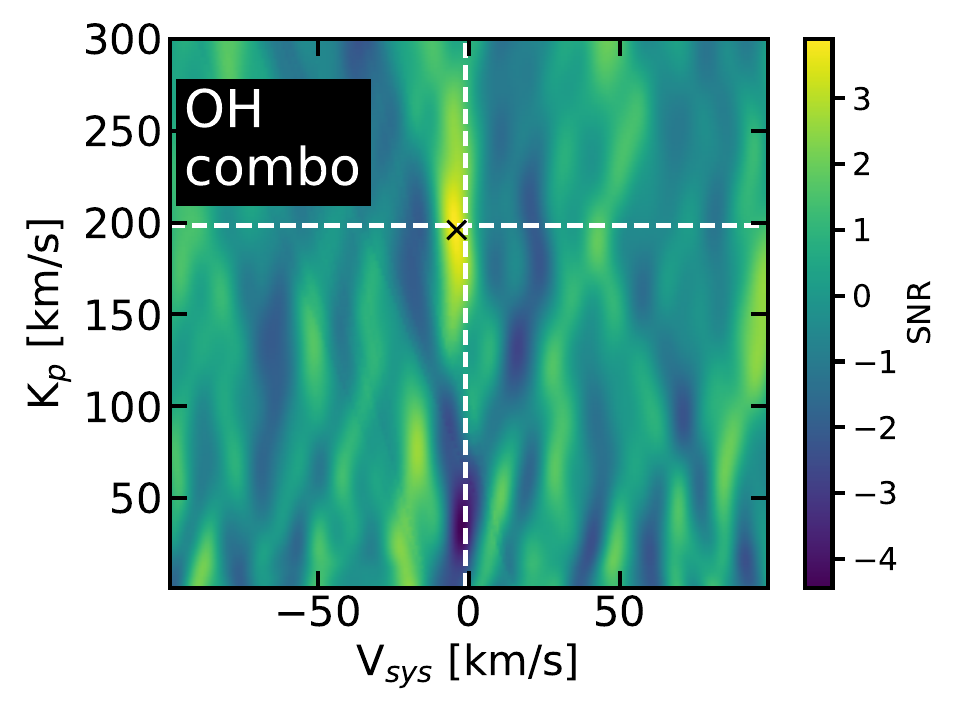}
    \includegraphics[width=0.24\linewidth]{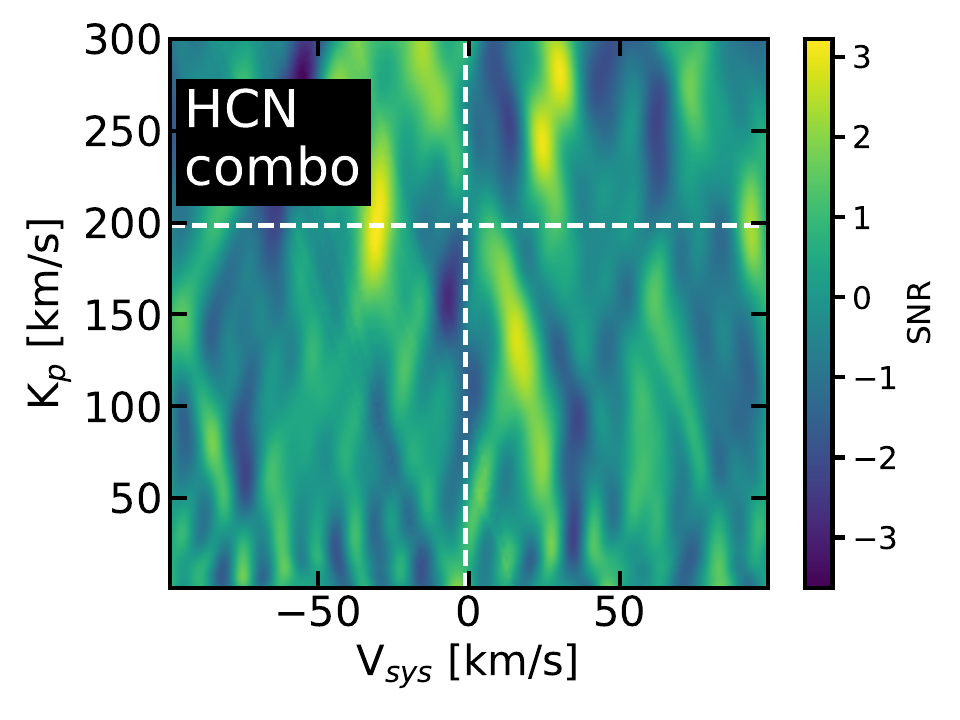}
    \caption{Cross-correlation signal-to-noise ratio (SNR) as a function of systemic velocity ($V_{sys}$) and planet Keplerian velocity ($K_{p}$) for detections of, from left to right, H$_{2}$O, CO, and OH on night 1 (top), night 2 (middle) and from both nights combined (bottom). SNR is calculated following Equation~\ref{eq:sigma} after performing a $3\sigma$ clipping using \texttt{astropy.stats.sigma\_clipped\_stats}. For the CO detection, we used only the orders containing the two distinct absorption bands near 1.61 and 2.45~$\mu$m. Black x marks denote the maximum cross-correlation in each plot, while white dashed lines indicate expected values for WASP-76b from previous studies \citep{West2016,Gaia2018}. All three molecules are clearly detected in both nights of data, and the combined data set shows SNRs of $6.93$, $6.47$, and $3.90$ for H$_{2}$O, CO, and OH, respectively. The right-most column shows non-detections of HCN on each night and in the combined data set.}
    \label{fig:CCFs}
\end{figure*}

Figure~\ref{fig:CCFs} shows a summary of our results. We clearly detected H$_{2}$O, CO, and OH, with the combined data set using both nights showing detection SNRs of $6.93$, $4.26$, and $3.90$, respectively. Across the wavelength range covered by IGRINS, the CO signature is dominated by two distinct absorption bands at 1.61 and 2.45~$\mu$m. We tested performing a cross-correlation for CO using only orders which cover these bands, and we found the SNR increased to $6.47$. H$_{2}$O and OH have broader absorption features covering most of the IGRINS wavelength range, so we used all orders to detect those molecules. We searched for a wide range of other gases, including CH$_{4}$, HCN, NH$_{3}$, SiO, TiO, VO, CaH, FeH, H$_{2}$S, and the isotope $^{13}$CO, but did not find any further significant detections.

\begin{figure*}
    \centering
    \includegraphics[width=0.32\linewidth]{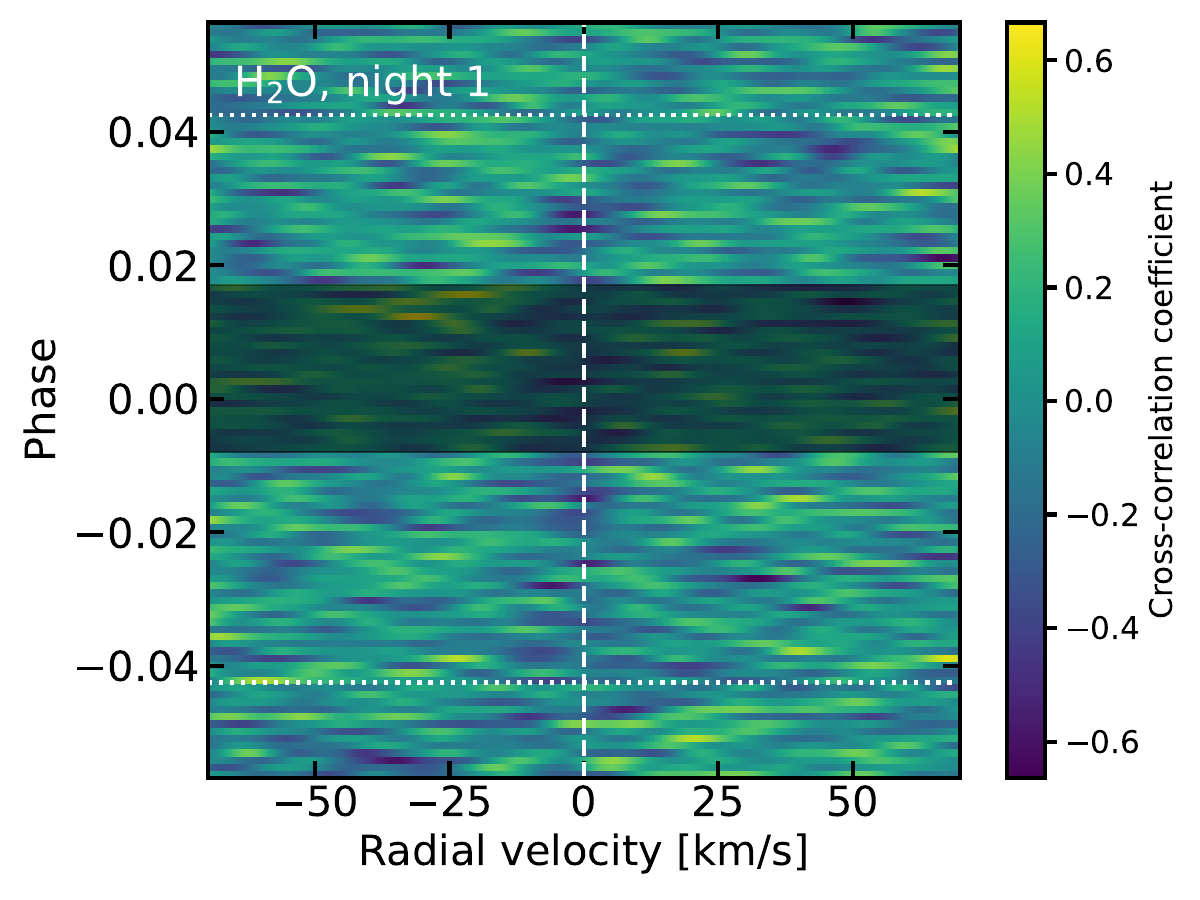}
    \includegraphics[width=0.32\linewidth]{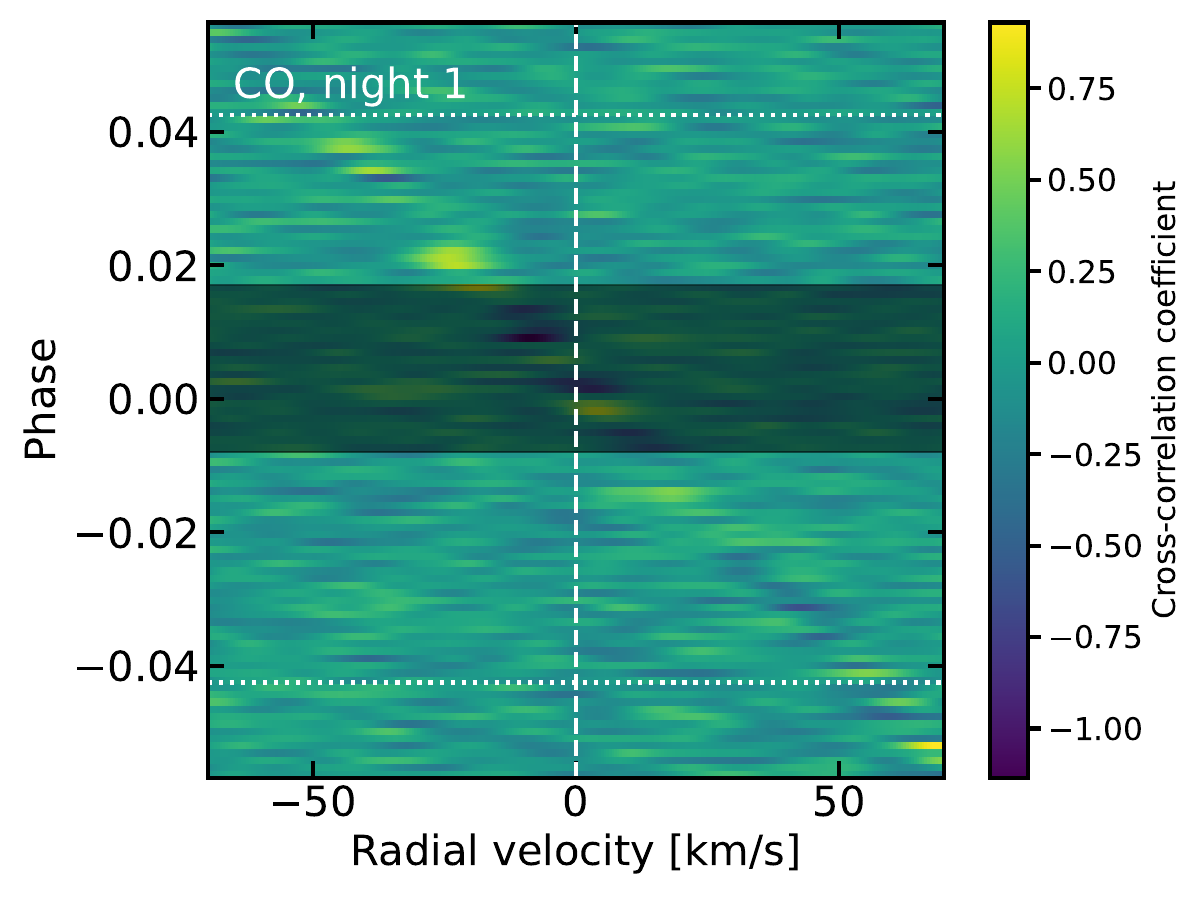}
    \includegraphics[width=0.32\linewidth]{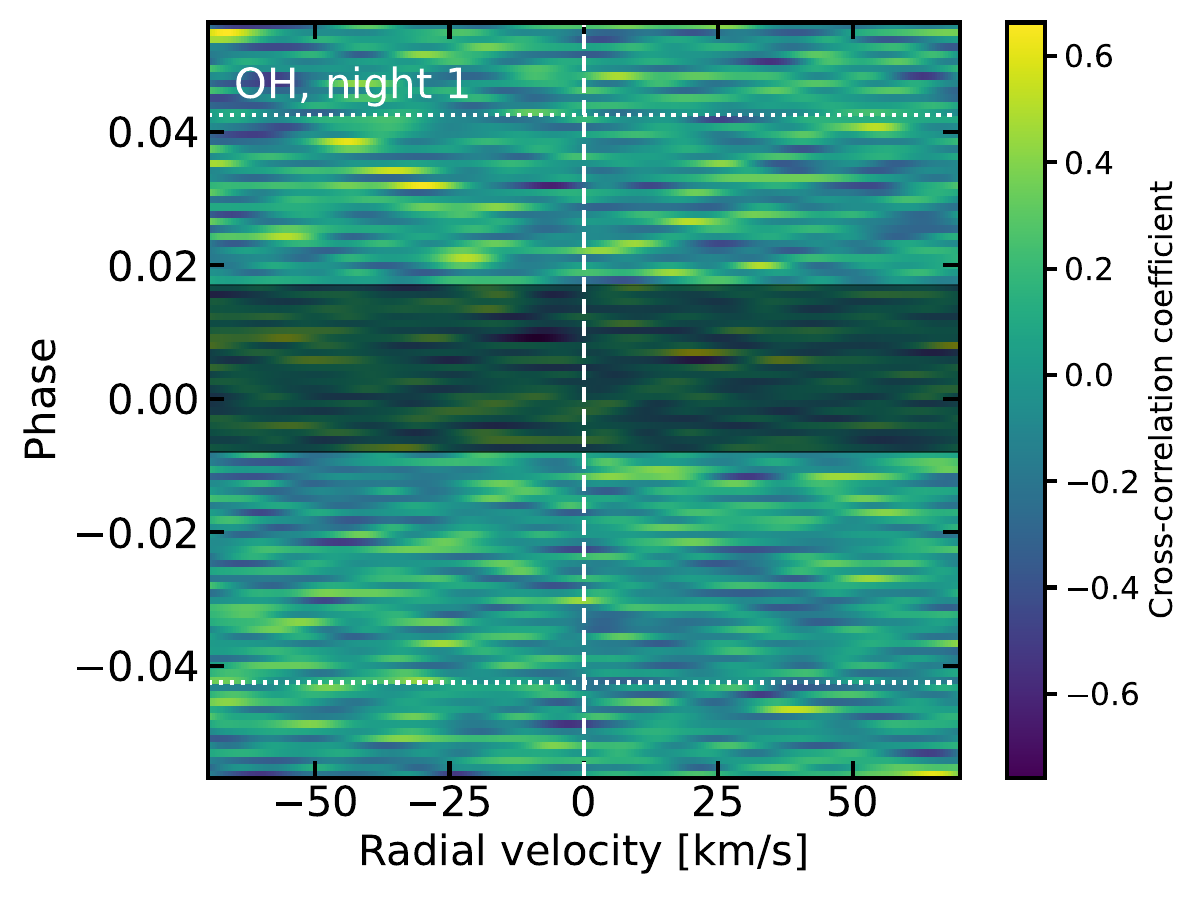}
    \includegraphics[width=0.32\linewidth]{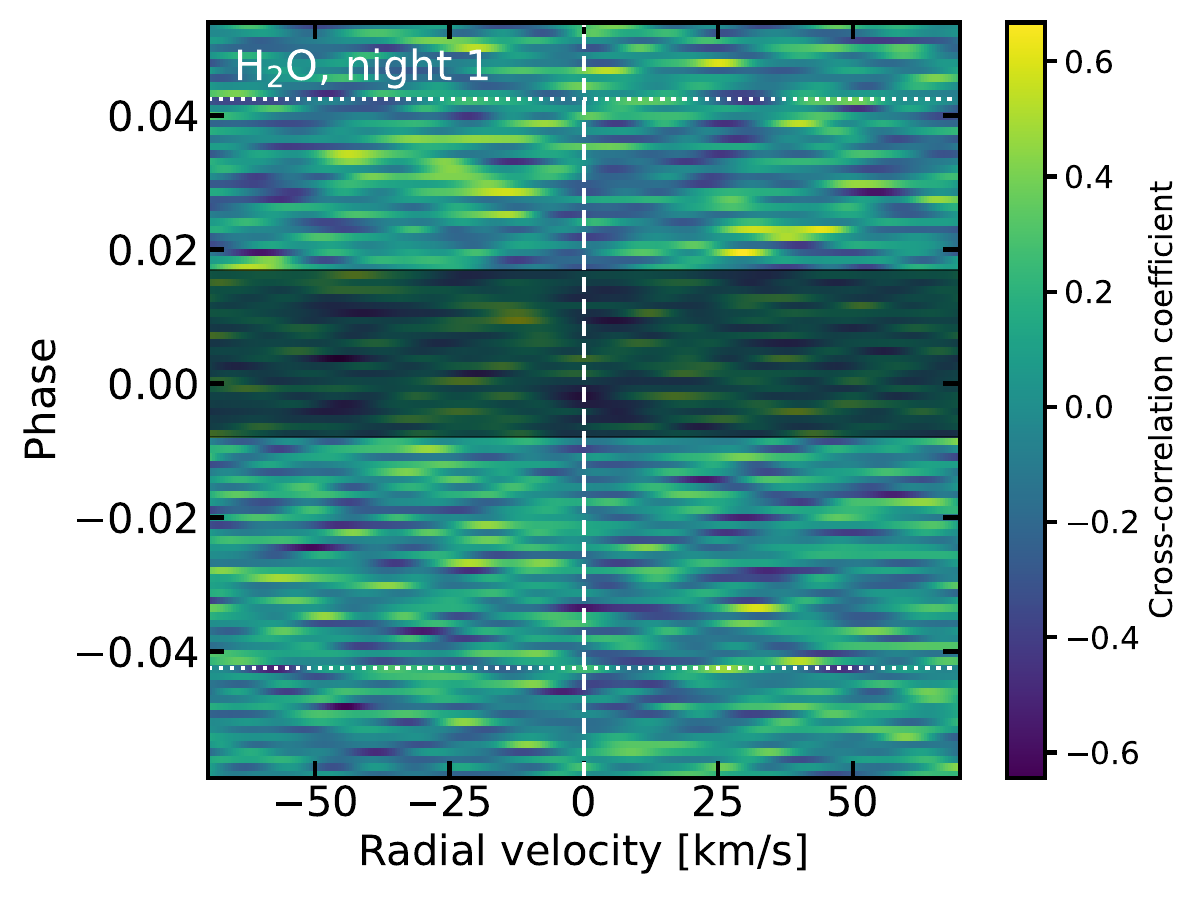}
    \includegraphics[width=0.32\linewidth]{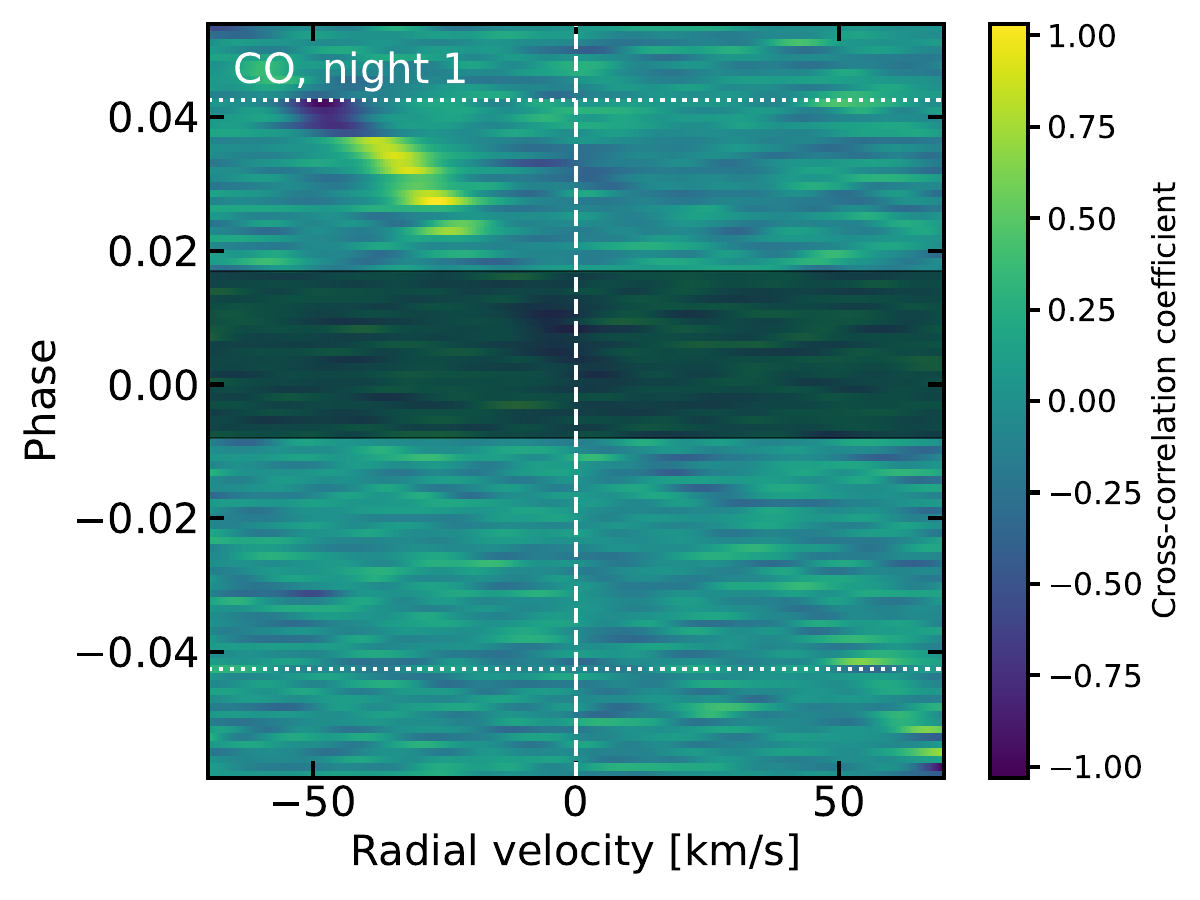}
    \includegraphics[width=0.32\linewidth]{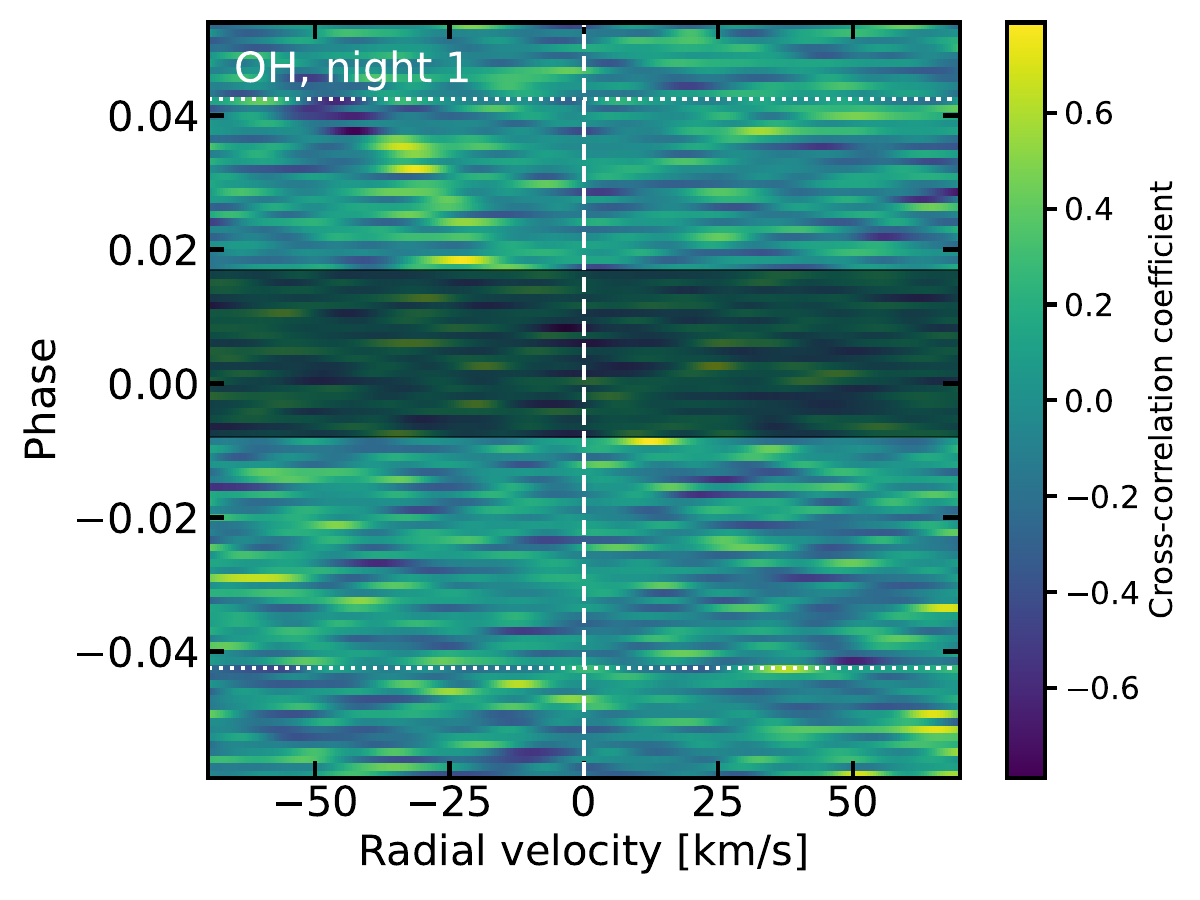}
    \caption{Trail plots showing cross-correlation strength as a function of radial velocity shift and orbital phase, in the rest frame of the planet WASP-76b, for H$_{2}$O (left), CO (middle), and OH (right) on night 1 (top) and night 2 (bottom). On all plots, horizontal white dotted lines indicate the beginning and end of transit and the vertical white dashed line indicates the location of the planetary signal. Signals in the trace that move diagonally from the upper left to lower right show the cross-correlation detections of the signal from the host star WASP-76 and tellurics. Grey shading indicates phases which were masked during the retrievals to remove any contamination from the stellar and telluric signals.}
    \label{fig:trail}
\end{figure*}

In addition to detecting CO and OH in the atmosphere of WASP-76b, we found residual CO and OH trails at constant, near-zero systemic velocity. We attribute the CO to a residual signal from the host star WASP-76 and the OH to residual telluric airglow \citep{Oliva2015}. The two signals appear to overlap in velocity because of the relatively small barycentric velocity during both observations (6.12~km/s and 4.38~km/s on night 1 and 2, respectively). 
To keep these stellar and telluric signals from influencing the derived planetary CO and OH abundances, 
we applied a mask to the data which ignored contributions to the cross-correlation strength from exposures at orbital phases $-0.008<\phi<0.017$ where the planetary trail crosses the same velocities as the stellar and telluric trails (see Figure~\ref{fig:trail}). The results we present here applied this mask evenly to all orders. However, we also tested applying this mask only to orders covering the strongest CO and OH features and found that 
the resulting elemental abundances, metallicity, and C/O ratio we derived for WASP-76b were within $1\sigma$ of the values retrieved from our main analysis.

We also identified a feature on night 2 where the strength of the cross-correlation with the stellar and/or telluric signals suddenly became stronger and then switched to a strong anti-correlation at orbital phases of $\phi>0.02$ (see Figure~\ref{fig:trail}). This feature is present in both the CO and OH trail plots. While the cause of this feature is unknown and outside the scope of this paper, it did not influence our conclusions because it showed no overlap with the planetary signal, as the planet's line-of-sight velocity at those phases is large enough to Doppler shift its signal to the point where it is clearly separated from the stellar signal.

\section{Retrieval Analysis}
\label{sec:retrieve}

In order to investigate the impact of different retrieval frameworks on our inferred abundances, we performed two types of chemistry retrievals: a free chemistry retrieval (Section~\ref{sec:chemret}), and a self-consistent grid-based retrieval (Section~\ref{sec:grid}). This choice is motivated by earlier work on WASP-18b \citep{Brogi2023}, where they found a different composition between free and self-consistent retrieval approaches. We additionally selected these two types of retrievals as representative end members of many possible types of retrievals, with one as free as possible and the other as self-consistent as possible.

\subsection{Free Retrievals of Chemical Abundances}
\label{sec:chemret}

We first fit the data with retrievals based on the log-likelihood framework developed in \citet{Brogi2019}, and using the model atmosphere framework 
described in \citet{Line2021}. We used the up-to-date line lists for H$_{2}$O, CO, and OH mentioned in Section~\ref{sec:ccor}, as well as  H$_{2}$-H$_{2}$ and H$_{2}$-He collision-induced absorption \citep{Karman2019}, and H$^{-}$ bound-free and free-free absorption \citep{Bell1987,John1988}.

We parameterized our one-dimensional atmosphere models with constant-with-altitude volume mixing ratios for the three detected gases and H$^{-}$ bound-free and free-free opacity ($n_{H_{2}O}$, $n_{CO}$, $n_{OH}$, $n_{HmBF}$, $n_{HmFF}$), a three-parameter Guillot temperature-pressure (T-P) profile \citep[$T_{0}$, $\gamma_{1}$, and $\kappa_{IR}$;][]{Guillot2010}, 
and a cloud-top pressure ($P_{c}$). We additionally fit for the planet Keplerian and system velocities ($K_{p}$ and $V_{sys}$), which affect the Doppler shift at which the planetary transmission signature is detected. Finally, we included as a nuisance parameter a scale factor on the reference planet radius ($\times R_{p}$). 
We initially fit for a phase offset to account for errors in the reported ephemeris, but we removed this from the final fits after initial retrievals showed it was consistent with zero. Our retrievals therefore contained a total of 12 free parameters.

Following \citet{Line2021}, we convolved the model spectra with a kernel for instrumental broadening. 
We also applied a kernel for planetary rotation broadening during transit, following Equation 15 from \citet{Gandhi2022} and assuming a rotational velocity of 5.14~km/s. We multiplied the model (1-$(R_p/R_*)^2$), appropriately Doppler shifted according to the Earth's barycentric velocity and the planet's $K_{p}$, $V_{sys}$, and orbital phase, by a matrix representing the eigenvectors discarded by the SVD detrending method we used to clean our data, creating a ``model-injected'' data cube. We then re-applied SVD detrending to this model injected data cube, cross-correlated the model with the SVD detrended data, and followed the methods of \citet{Brogi2019} to convert the cross-correlation strength to a log-likelihood. We then evaluated the log-likelihood within the context of the Python package \texttt{pymultinest} \citep{Buchnerpymulti} to perform Bayesian inference and model selection. The only difference between our methods and those of \citet{Line2021} is that, as our observation is a transmission spectrum, our models are in units of transit depth, or the planet-to-star radius ratio squared (($R_{p}$/$R_{*}$)$^{2}$).

\begin{figure*}
    \centering
    \includegraphics[width=\linewidth]{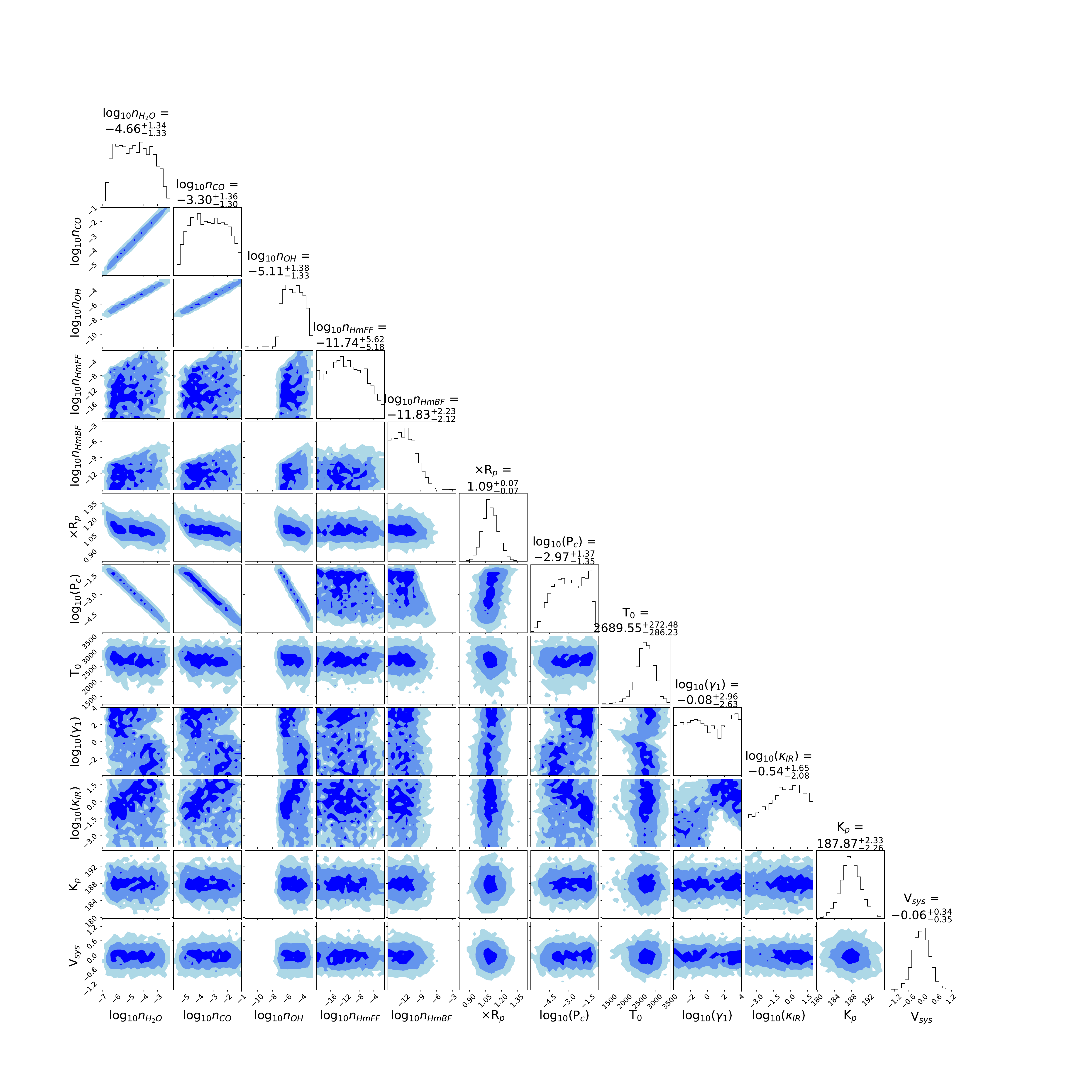}
    \caption{Posterior distributions for all parameters in the fiducial retrieval. Off-diagonal plots show 2D posterior probabilities for pairs of parameters, with 1, 2, and $3\sigma$ intervals shaded in dark, medium, and light blue. On-diagonal plots show marginalized posterior probability distributions for each parameter.}
    \label{fig:pairs_fiducial}
\end{figure*}

Figure~\ref{fig:pairs_fiducial} shows the constraints derived from this retrieval. We retrieved abundances of $\log_{10}(n)=-4.66^{+1.34}_{-1.33}$, $-3.30^{+1.36}_{-1.30}$, and $-5.11^{+1.39}_{-1.33}$ for H$_{2}$O, CO, and OH, respectively. 
We converted these abundances into a total volatile metallicity ([(C+O)/H]) and carbon-to-oxygen ratio (C/O) relative to solar values using the equations
\begin{multline}
    \left[\frac{\mathrm{C}+\mathrm{O}}{\mathrm{H}}\right]=\log_{10}\left(\frac{2n_{\mathrm{CO}}+n_{\mathrm{H}_{2}\mathrm{O}}+n_{\mathrm{OH}}}{2n_{\mathrm{H}_{2}}}\right)\\-\log_{10}\left((\mathrm{C}/\mathrm{H})_{\odot}+(\mathrm{O}/\mathrm{H})_{\odot}\right),
\end{multline}
and
\begin{equation}
    \mathrm{C}/\mathrm{O}=\frac{n_{\mathrm{CO}}}{n_{\mathrm{H}_{2}\mathrm{O}}+n_{\mathrm{CO}}+n_{\mathrm{OH}}},
\end{equation}
respectively. We used the solar C and O abundances as references because, while WASP-76 has a measured iron abundance of [Fe/H]$=0.23\pm0.10$, it has no measured carbon or oxygen abundances \citep{West2016}. Additionally, prior research has shown that solar-type stars generally have near-solar C/O ratios \citep{Fortney2012,Bedell2018}. We refer throughout the rest of the paper to WASP-76b's [Fe/H] when comparing metallicity values but to the solar C/O when comparing C/O ratios. 
We derived a volatile metallicity of $\left[\frac{\mathrm{C+O}}{\mathrm{H}}\right]=-0.14^{+1.36}_{-1.30}$ and a C/O ratio of $\mathrm{C/O}=0.94\pm0.02$.

\begin{figure}
    \centering
    \includegraphics[width=\linewidth]{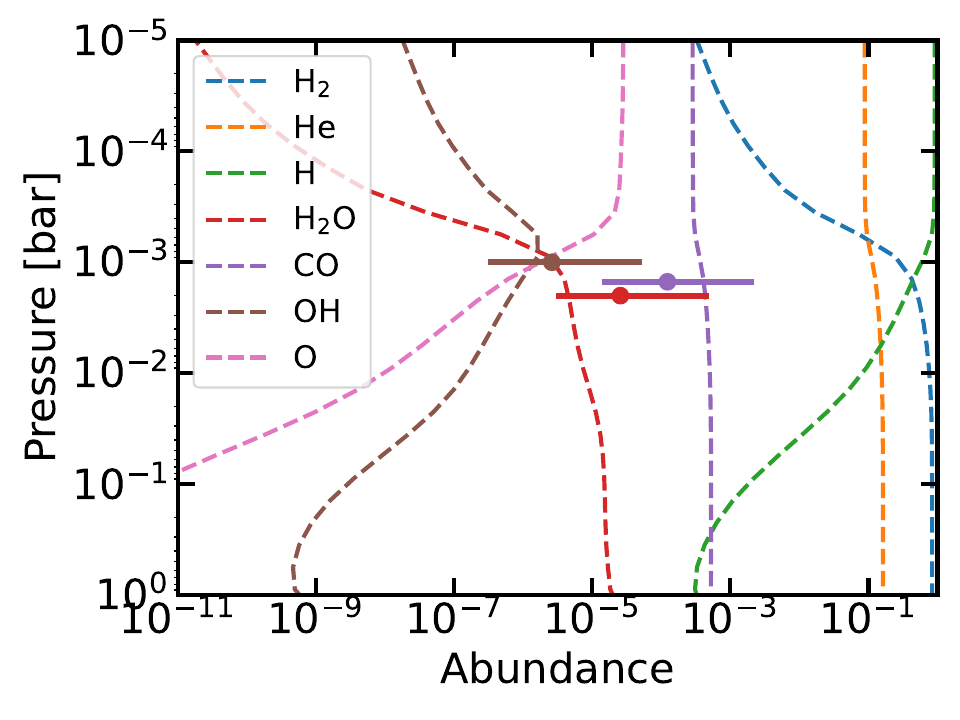}
    \caption{Comparison of retrieved abundances from the fiducial retrieval to expectations for a solar composition gas in chemical equilibrium. Dashed lines show expected abundance of various species as a function of pressure for a model with a composition consistent with the best-fit results from this retrieval ([M/H]$=-0.125$ and C/O$=0.9$). Solid points with $1\sigma$ error bars show retrieved abundances for H$_{2}$O (red), CO (purple), and OH (brown). The points are placed near the mean photospheric pressure across the observed wavelengths.}
    \label{fig:equcomp_fiducial}
\end{figure}

At face value, these results suggest that WASP-76b has a super-stellar metallicity but a significantly super-solar C/O ratio. However, at the high temperatures achieved in the upper atmosphere of WASP-76b, water dissociation is predicted to influence the retrieved oxygen abundance \citep[e.g.,][]{Parmentier2018}. Figure~\ref{fig:equcomp_fiducial} compares our retrieved abundances to expectations from an equilibrium chemistry model for WASP-76b at a solar metallicity and C/O ratio. Above pressures of $\approx10^{-2}$~bar, water is expected to dissociate into both OH molecules and atomic oxygen. Our data revealed the presence of both H$_{2}$O and OH, confirming that dissociation is influencing our observations. Additionally, a tentative detection of atomic O has been reported for WASP-76b in optical MAROON-X observations \citep{Pelletier2023}. IGRINS is not sensitive to atomic O, but at solar composition, chemical equilibrium models predict that $\lesssim3$\% of the oxygen would be in atomic form at the photosphere ($\approx10^{-2}-10^{-3}$~bar), which will impact our estimates of the metallicity and C/O ratio by an amount much smaller than our retrieved error bars. The negligible influence of unaccounted atomic O differs significantly from the case of \citet{Brogi2023}, which studied the higher-temperature, higher-gravity planet WASP-18b and found that $20-30$\% of oxygen was in atomic form due to thermal dissociation, thus significantly influencing the estimated C/O ratio. 

However, equilibrium chemistry models also indicate a large change in the abundance of water with pressure, due to the higher upper atmosphere temperatures driving dissociation of water, as seen in Figure~\ref{fig:equcomp_fiducial}. 
Our models which assume a constant abundance with pressure may therefore bias the retrieved abundance. To determine whether this bias shapes our retrieved volatile metallicity and C/O ratio, we ran a second set of retrievals where each molecular abundance profile was represented by two parameters: a deep atmosphere abundance ($n_{\mathrm{H}_{2}\mathrm{O}}$, $n_{\mathrm{CO}}$, and $n_{\mathrm{OH}}$) and a break pressure above which the abundance drops to zero ($P_{b,\mathrm{H}_{2}\mathrm{O}}$, $P_{b,\mathrm{CO}}$, and $P_{b,\mathrm{OH}}$).

\begin{figure*}
    \centering
    \includegraphics[width=\linewidth]{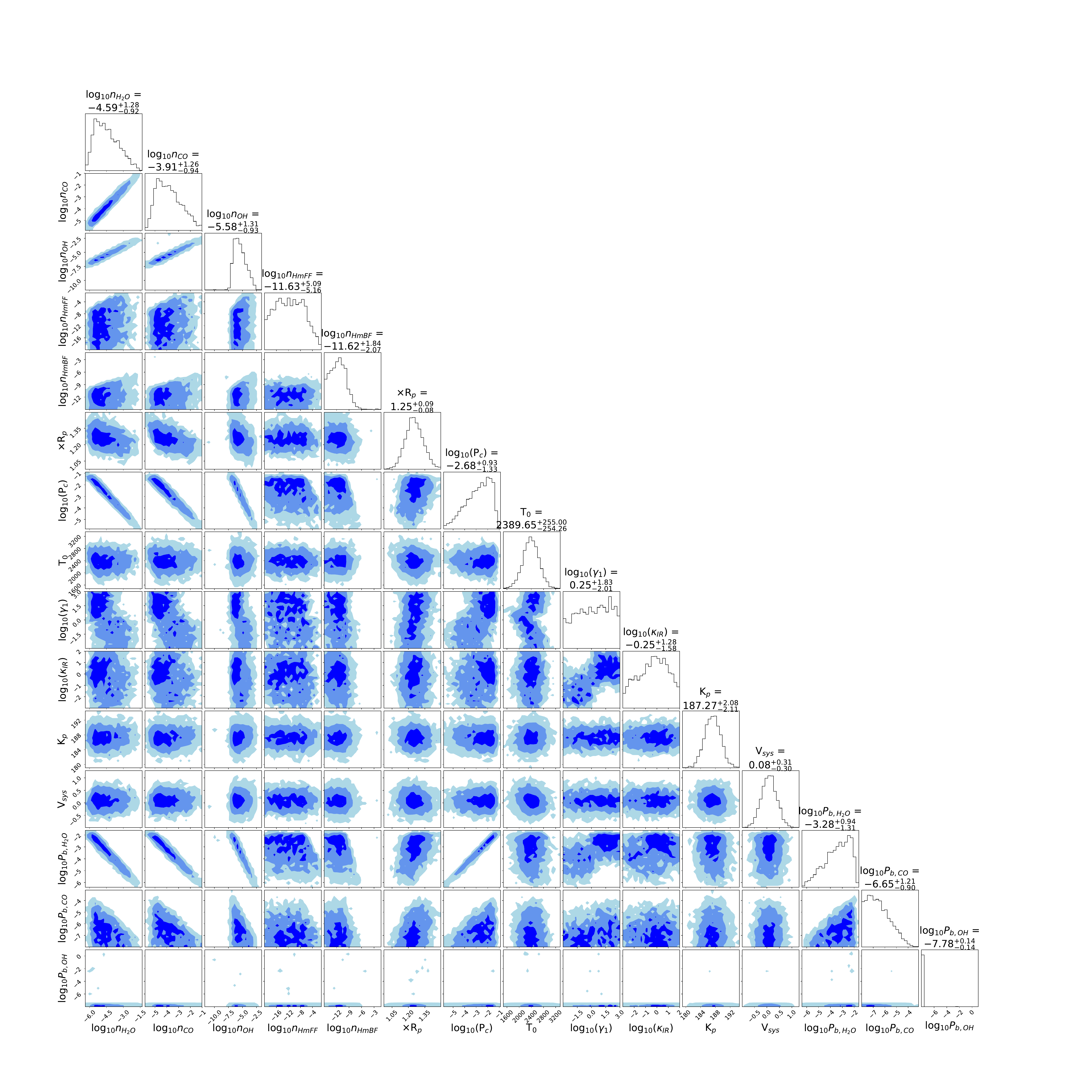}
    \caption{Same as Figure~\ref{fig:pairs_fiducial} but for the break pressure retrieval. In this retrieval, each detected molecule is represented with two parameters: a deep atmosphere abundance ($\log_{10}(n)$) and a break pressure where the abundance drops to zero ($P_{b}$).}
    \label{fig:pairs_breakp}
\end{figure*}

\begin{figure}
    \centering
    \includegraphics[width=\linewidth]{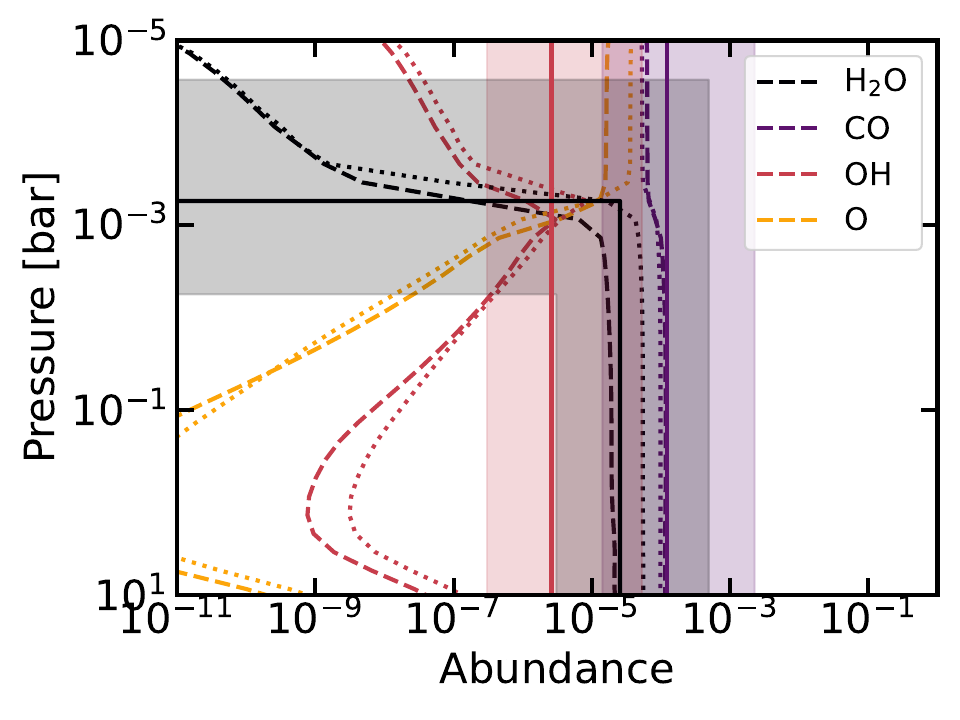}
    \caption{Comparison of retrieved abundances from the break pressure retrieval to expectations for models in chemical equilibrium. Solid lines with $1\sigma$ shaded error bars show retrieved profiles for H$_{2}$O (black), CO (purple), and OH (red). As described in Section~\ref{sec:chemret}, each abundance was fit with a two-parameter model including a deep atmosphere abundance and a pressure at which the abundance dropped to 0. For CO and OH, the retrieved break pressure was consistent with the top of the atmosphere, indicating the retrieval preferred a model with constant abundances with altitude for both gases. Dashed and dotted lines show predictions from chemical equilibrium models with compositions consistent with the best-fit results from this retrieval ([M/H]$=-0.75$ and C/O$=0.8$) and from the self-consistent gridtrieval ([M/H]$=-0.75$ and C/O$=0.6$), respectively.}
    \label{fig:equcomp_breakp}
\end{figure}

Figures~\ref{fig:pairs_breakp} and \ref{fig:equcomp_breakp} shows the results of this pressure break-point retrieval. The retrieved deep atmosphere abundances of H$_{2}$O, CO, and OH are $\log_{10}(n)=-4.59^{+1.28}_{-0.92}$, $-3.91^{+1.26}_{-0.94}$, and $-5.59^{+1.31}_{-0.93}$, respectively. 
Based on these deep atmosphere abundances, the retrieved volatile metallicity and carbon-to-oxygen ratio are $\left[\frac{\mathrm{C}+\mathrm{O}}{\mathrm{H}}\right]=-0.70^{+1.27}_{-0.93}$ and C/O$=0.80^{+0.07}_{-0.11}$. 
The retrieved break pressures for CO and OH are consistent with the top of the atmosphere in the models, which indicates that the retrieval preferred a model where the CO and OH abundances are constant with altitude.\footnote{We note that, based on Figure~\ref{fig:equcomp_breakp}, a deep atmosphere abundance and break pressure are not a perfect model for OH, which is expected to increase with altitude until a break point and then decrease again. We used this simplified model because our OH detection was not strong enough to constrain a more complex model with more parameters, and it is possible that a more complex model would show an OH abundance profile more similar to the model prediction.} However, the retrieved break pressure for H$_{2}$O is $\log_{10}P_{b,\mathrm{H}_{2}\mathrm{O}}=-3.28^{+0.94}_{-1.31}$. 
As shown in Figure~\ref{fig:equcomp_breakp}, this is consistent with the expected pressure at which the water abundance would start to significantly decrease due to dissociation in an atmosphere in equilibrium and with approximately the same metallicity and C/O ratio as what was retrieved.

We also investigated performing retrievals separately on the first and second half of the transits, to search for inhomogeneities between the two limbs. However, our data were not high enough signal-to-noise to constrain any limb asymmetries.

\subsection{Self-Consistent Gridtrievals}
\label{sec:grid}

\begin{figure*}
    \centering
    \includegraphics[width=0.8\linewidth]{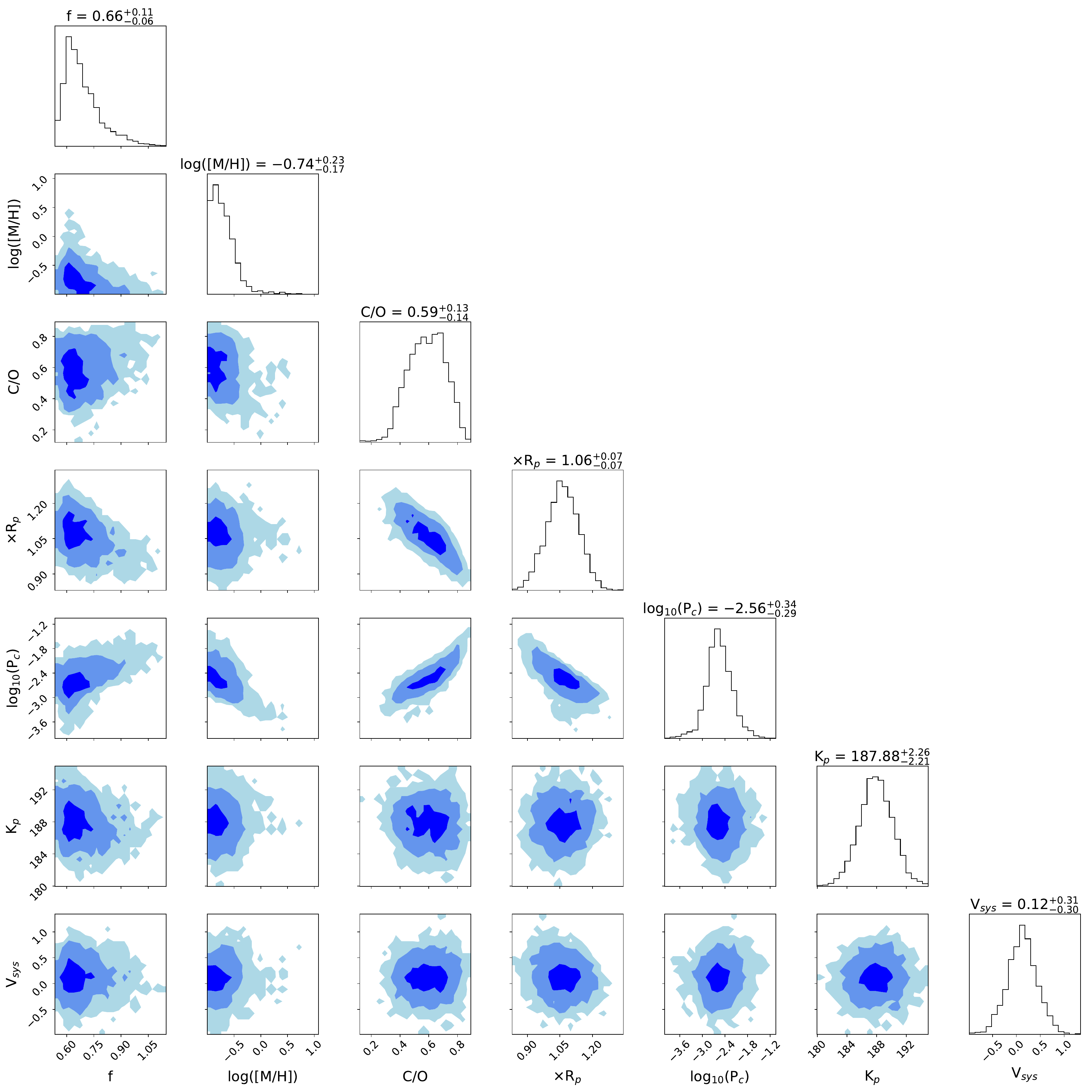}
    \caption{Same as for Figure~\ref{fig:pairs_fiducial}, but for the gridtrieval described in Section~\ref{sec:grid}.}
    \label{fig:gridpairs}
\end{figure*}

In addition to the free chemistry retrievals listed above, we ran a self-consistent ``gridtrieval'' for WASP-76b using the grid-based retrieval framework described in \citet{Brogi2023}. The gridtrieval framework fits the data by interpolating between pre-computed models which self-consistently calculate the T-P profile and molecular chemistry based on provided elemental abundances. We used a framework identical to that described in \citet{Brogi2023} to calculate the grid of self-consistent models for WASP-76b. The gridtrieval parameterizes the atmospheric composition and temperature with three parameters: a heat redistribution efficiency parameter ($f$), an atmospheric metallicity ([M/H]) which scaled all elements (re-normalizing to H), and a carbon-to-oxygen ratio (C/O) (adjusted while preserving the sum of C+O after the [M/H] scaling). In addition to these three parameters, we also fit for a radius scale factor ($\times R_{p}$), cloud-top pressure ($P_{c}$), and velocities ($K_{p}$ and $V_{sys}$) as described above, for a total of seven free parameters. A pairs plot for the gridtrieval is shown in Figure~\ref{fig:gridpairs}.

The gridtrieval retrieved a metallicity of [M/H]$=-0.74^{+0.23}_{-0.17}$ 
and a carbon-to-oxygen ratio of C/O$=0.59^{+0.13}_{-0.14}$. 
The expected atmospheric composition for a planet in chemical equilirium with this metallicity and C/O ratio are also shown in Figure~\ref{fig:equcomp_breakp}. These values are consistent to within $1\sigma$ with the result derived from the break pressure retrieval. We note that grid-based retrievals tend to produce tighter constraints than more flexible free retrievals owing to the assumption of 1D radiative-convective-thermochemical equilibrium which rules out various abundance/temperature combinations that don't fall within those self-consistent assumptions \citep[e.g.,][]{Brogi2023}.

\section{Retrievals of Velocity Offsets}
\label{sec:velret}

Previous observations of WASP-76b at optical wavelengths have shown the ``Ehrenreich effect'', or an asymmetry in the velocity at which iron \citep{Ehrenreich2020} and other atomic species \citep{Pelletier2023} were detected as a function of phase. This asymmetry has been suggested to be due to differences in Fe abundance \citep{Ehrenreich2020}, temperature-pressure (T-P) profiles \citep{Wardenier2021,Pelletier2023}, cloud opacity \citep{Savel2022,Pelletier2023}, and/or wind speeds \citep{Gandhi2022} between the eastern and western limbs. 
As shown in Figure~\ref{fig:trail}, our trail plots do not show high enough signals to detect deviations in radial velocity for 
H$_{2}$O, CO, or OH as a function of orbital phase as was found for iron. However, Figure~\ref{fig:CCFs} shows that on both nights we see differences in the overall $V_{sys}$ and $K_{p}$ at which each of these three species are detected. 
In order to investigate the significance of these velocity differences, we performed another set of atmospheric retrievals where the only free parameters were $V_{sys}$, $K_{p}$, and $\times R_{p}$. We did one retrieval each for the three gases whose abundances we constrain, and we fixed the abundance profile of the gas 
to the best-fit values from the break pressure retrieval. Table~\ref{tab:kpvsys} lists the offsets between the retrieved velocities and the values expected from previous radial velocity observations of WASP-76b's orbit \citep[$K_{p}=196.52\pm0.94$~km/s and $V_{sys}=-1.11\pm0.50$~km/s,][]{Ehrenreich2020,Gaia2018}. We compared these velocity offsets to those predicted from 3D general circulation models of WASP-76b \citep{Wardenier2023} to interpret them in terms of the expected circulation patterns in the planet's atmosphere.

\begin{table}
\centering
    \begin{tabular}{c | c c}
    \textbf{Gas} & \textbf{$\Delta K_{p}$ [km/s]} & \textbf{$\Delta V_{sys}$ [km/s]} \\
    \hline
    H$_{2}$O & $-1.1^{+3.0}_{-2.9}$ & $2.42^{+0.39}_{-0.38}$ \\
    CO & $-18.8\pm2.7$ & $0.69\pm0.38$ \\
    OH & $3.2^{+5.5}_{-5.2}$ & $-2.97^{+0.70}_{-0.75}$ \\
    \end{tabular}
    \caption{\label{tab:kpvsys} Retrieved systemic velocity ($V_{sys}$) and Keplerian velocity ($K_{p}$) offsets from expected values \citep{West2016,Gaia2018} for retrievals where only velocity offsets and a scale factor were allowed to vary.}
\end{table}

\citet{Wardenier2023} divide atmospheric species into those whose signal primarily comes from the dayside or nightside of the planet. H$_{2}$O is classified as a nightside species, as dissociation on the dayside reduces the H$_{2}$O abundance in that hemisphere. Conversely, dayside dissociation increases the OH abundance, so OH is classified as a dayside species. Finally, while CO is expected to be present across the entire planet in similar abundances \citep{Savel2023}, the hotter temperature on the dayside means that the majority of the CO absorption originates from the dayside, making it a dayside species.

In their 3D models, dayside species exhibit a negative shift in both $V_{sys}$, because planetary winds overall blueshift the absorption features, and $K_{p}$, because in 3D models the signals become more blueshifted throughout the course of the transit \citep{Wardenier2023}. While our data are not precise enough to resolve an increasing blueshift in the trail plots shown in Figure~\ref{fig:trail}, we do find a negative $K_{p}$ shift for CO of $\Delta K_{p}=-18.8$~km/s. This is consistent with the findings of \citet{Wardenier2023}, who found a $\Delta K_{p}\approx20$ km/s for CO in most of their models. For OH, we do not see a similar shift in $K_{p}$, as predicted by \citet{Wardenier2023}. However, the relative offset in $V_{sys}$ between CO and OH is similar to what is found by \citet{Wardenier2023} - their models predict OH to have a $V_{sys}$ about $0.5-1$~km/s less than that of CO, while we find OH to have a $V_{sys}$ 3.6~km/s less than CO.

For the nightside species H$_{2}$O, our slight $K_{p}$ offset of $-1.1$~km/s is consistent with the results of \citet{Wardenier2023}, which predict a smaller offset of about $5-10$~km/s for nightside species. However, the redshifted $V_{sys}$ we see for H$_{2}$O cannot be explained by any of their models. In theory, H$_{2}$O could have a redshifted $V_{sys}$ on a windless planet if it was present on the colder leading limb but dissociated on the warmer trailing limb, because in this scenario the H$_{2}$O detection would come entirely from a region of the planet rotating away from the observer \citep{Wardenier2023}. However, our data are not precise enough to detect phase-resolved differences in abundances, so we cannot confirm whether H$_{2}$O is indeed only present on the leading limb. In addition, when realistic wind speeds are added to the planet, none of the models of \citet{Wardenier2023} show a net redshifted $V_{sys}$ for H$_{2}$O. However, models incorporating magnetic drag do retrieve similar velocity differences as what we see between H$_{2}$O and CO, even if the absolute values are not as redshifted \citep{Beltz2023}. More precise phase-resolved data or models incorporating more effects such as different magnetic drag parameterizations \citep{Beltz2023} will be required to fully understand our detection of redshifted $V_{sys}$ for H$_{2}$O.

\section{Discussion}
\label{sec:discuss}

\subsection{Comparison to Previous Carbon and Oxygen Detections}
\label{sec:prevdetect}
Several of the species we detected have been previously measured in transmission spectra of WASP-76b. H$_{2}$O has been previously detected using \textit{HST} \citep{Edwards2020,Fu2021}. The two published \textit{HST} retrievals disagree on the overall abundance of water - \citet{Edwards2020} reported a retrieved H$_{2}$O abundance of $\log_{10}(n_{\mathrm{H}_{2}\mathrm{O}})=-2.85^{+0.42}_{-0.71}$, while \citet{Fu2021} found an oxygen abundance of $\log_{10}(n_{\mathrm{O}})=-4.34^{+0.43}_{-0.48}$. In the break pressure model, we found an overall oxygen abundance of $\log_{10}(n_{\mathrm{O}})=\log_{10}(n_{\mathrm{H}_{2}\mathrm{O}}+n_{\mathrm{OH}}+n_{\mathrm{CO}})=-3.82^{+1.05}_{-0.78}$, which is consistent with the results of both \citet{Edwards2020} and \citet{Fu2021}.

Dayside observations of WASP-76b both at high-resolution with CRIRES+ \citep{Yan2023} and at low-resolution with \textit{Spitzer} also showed detections of CO in its atmosphere. \citet{Yan2023} found a CO abundance of $\log_{10}(n_{CO})=-3.6^{+1.8}_{-1.6}$, and \citet{Fu2021} report an overall carbon abundance of $\log_{10}(n_{C})=-4.42^{+0.48}_{-0.66}$. Our retrieved CO abundance is also consistent with both of these results within $1\sigma$. 

Previous high-resolution transit observations of WASP-76b have reported results in disagreement with each other. \citet{SanchezLopez2022} observed WASP-76b with CARMENES and reported a similar detection of H$_{2}$O, but at a significantly higher $K_{p}=249\pm38$ and lower $V_{sys}=-14.3\pm2.6$. They additionally reported a detection of HCN. On the other hand, \citep{Hood2024} observed WASP-76b with SPIRou and detected H$_{2}$O and CO but not HCN or OH. They reported best-fit abundances of $\log_{10}(n_{\mathrm{H}_{2}\mathrm{O}})=-4.52\pm0.77$ and $\log_{10}(n_{\mathrm{CO}})=-3.09\pm1.05$, with upper limits of $\log_{10}(n_{\mathrm{OH}})<-6$ and $\log_{10}(n_{\mathrm{HCN}})<-5.5$ and a resulting retrieved C/O ratio of $0.94\pm0.39$.

To compare our results to these previous high-resolution transit observations, we performed an additional retrieval where we added HCN as a free parameter. We found an unconstrained abundance with a $2\sigma$ upper limit of $\log(\mathrm{HCN})=-4.95$. Our results are thus in good agreement with those of \citet{Hood2024}. Our retrieved OH abundance and upper limit on the HCN abundance are consistent with their derived upper limits. Additionally, our H$_{2}$O and CO abundances and C/O ratio are within their $1\sigma$ errors. However, our results disagree with those of \citet{SanchezLopez2022}, both in our inability to detect HCN and in the different velocity shifts we retrieve for H$_{2}$O. We hypothesize that this may be due to a difference in data reduction methods. Both our reduction and that of \citet{Hood2024} used similar applications of principal component analysis/SVD, with the same number of components removed from all orders and across all molecules being analyzed. However, \citet{SanchezLopez2022} removed different numbers of components in order to optimize detections of each molecule individually. This may have resulted in a sporadic detection of HCN through over-optimization of the data reduction.

\subsection{Potential Formation Scenarios for WASP-76b and Exoplanet C/O Ratios in General}
\label{sec:formation}

Our results indicate that the atmosphere of WASP-76b has either a stellar metallicity and a super-solar C/O ratio (from the break pressure free retrieval) or a sub-stellar metallicity and a solar C/O ratio (from the gridtrieval). 
Previous research has suggested that a C/O ratio near consistent with solar and stellar or sub-stellar metallicities could be consistent with many origin locations within the disk, depending on factors such as the relative rate of solid and gas accretion \citep[e.g.,][]{Madhu2014,Khorshid2022}. On the other hand, a super-solar C/O ratio could result from gas-dominated accretion beyond the CO snowline in the protoplanetary disk and subsequent migration \citep[e.g.,][]{Madhu2014,Oberg2016,Madhu2017} or in situ formation between the soot line and water snow line \citep{Chachan2023}. These two scenarios of formation could be distinguished by the refractory element enrichment, as formation outside the CO snow line is predicted to show refractory enrichments $\lesssim1\times$ solar, while formation between the soot line and water snow line is expected to have higher refractory abundances of $\gtrsim10\times$ solar \citep{Chachan2023}.

\citet{Pelletier2023} and \citet{Gandhi2023} recently reported the abundances of the refractory elements Fe, Mg, Cr, Mn, Ni, V, Ba, and Ca on WASP-76b, which ranged from 0.14 to 18.20$\times$ solar. The intermediate values of these refractory abundances, with most between $1-10\times$ solar, make it difficult to distinguish between the two proposed formation scenarios. However, future work to measure abundances of other elements such as nitrogen \citep{Oberg2016,Ohno2023} and sulfur \citep{Polman2023} may provide more paths to understand the formation of WASP-76b. 

\begin{figure}
    \centering
    \includegraphics[width=\linewidth]{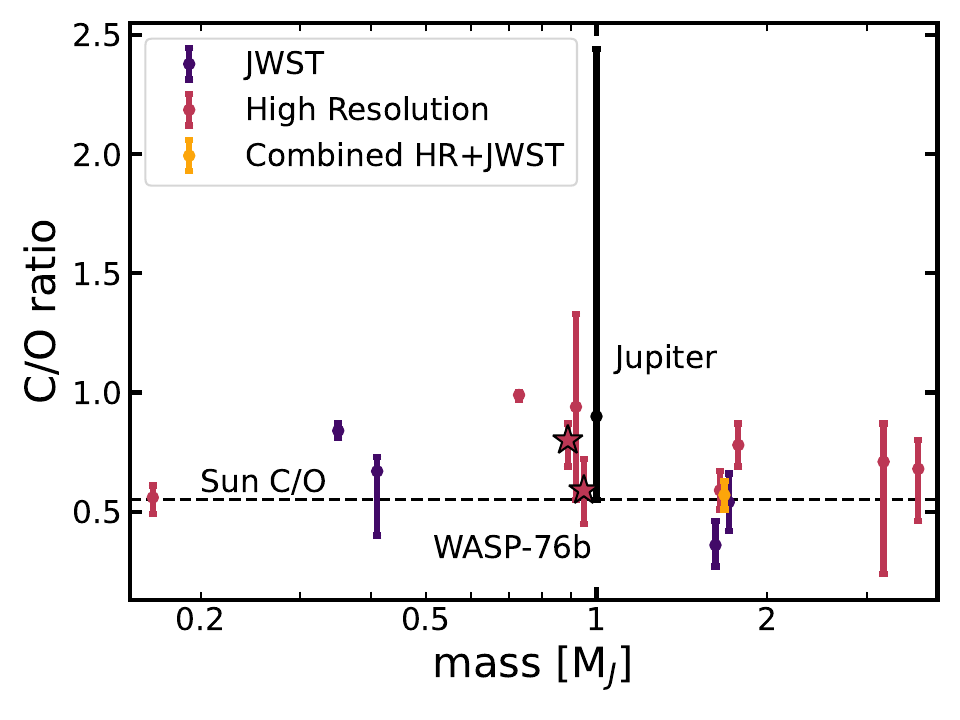}
    \caption{Carbon-to-oxygen (C/O) ratio as a function of mass for all planets with constrained abundances of both oxygen-bearing and carbon-bearing species. Purple, pink, and yellow points indicate results from \textit{JWST}, high-resolution observations, and combined analyses of both \textit{JWST} and high-resolution data (citations given in Table~\ref{tab:ctoo}). The black point indicates the value for Jupiter based on the Galileo and Juno missions. The pink points with black outlined, star-shaped markers are our results for WASP-76b. Planets with multiple measurements have small offsets in mass applied between the different results for clarity. While there is no clear trend with mass, overall the population of exoplanets shows a mean C/O ratio greater than solar.}
    \label{fig:CO_trend}
\end{figure}

While the C/O ratio of WASP-76b alone is inconclusive in determining its formation, the launch of \textit{JWST} in 2021 and the advent of methods for retrieving abundances from high-resolution spectroscopy are for the first time allowing intercomparison of a sample of exoplanets with C/O ratios constrained from observation of both carbon- and oxygen-bearing species. Figure~\ref{fig:CO_trend} and Table~\ref{tab:ctoo} shows the full sample of planets with constrained C/O ratios. While there are a wide variety of reported values, most are weighted toward super-solar C/O ratios. After calculating a single averaged value for each planet for which there are multiple reported values from different sources, the full sample has a weighted mean C/O ratio of 0.89 and a median of 0.72.

There is already significant scatter in the reported C/O values, indicating that there may be a variety of formation mechanisms at play. The sample of planets with well-constrained C/O ratios is also still relatively small, so future work to measure C/O ratios on a wider population of planets may further elucidate any population-level trends.

\begin{table*}[]
    \centering
    \begin{tabular}{c|c | c | c}
      Planet Name   & C/O Ratio & Provenance & Reference \\
      \hline \hline
      \multirow{2}{*}{HD 149026b} & $0.84\pm0.03$ & \multirow{2}{*}{\textit{JWST} (NIRCam)} & \citet{Bean2023} \\
      & $0.67^{+0.06}_{-0.27}$ & & \citet{Gagnebin2024} \\
      \hline
      HD 209458b & $0.99^{+0.01}_{-0.02}$ & High-res (CRIRES) & \citet{Gandhi2019} \\
      \hline
      HIP 65Ab & $0.71^{+0.16}_{-0.47}$ & High-res (IGRINS) & \citet{Bazinet2024} \\
      \hline
      MASCARA-1b & $0.68^{+0.12}_{-0.22}$ & High-res (CRIRES+) & \citet{Ramkumar2023} \\
      \hline
      WASP-43b & $0.78 \pm 0.09$ & High-res (CRIRES+) & \citet{Lesjak2023} \\
      \hline
      \multirow{3}{*}{WASP-76b} & $0.80^{+0.07}_{-0.11}$ & \multirow{2}{*}{High-res (IGRINS)} & \multirow{2}{*}{This work} \\
      & $0.59^{+0.13}_{-0.14}$ & & \\
      & $0.94\pm0.39$ & High-res (SPIRou) & \citet{Hood2024} \\
      \hline
      \multirow{4}{*}{WASP-77Ab} & $0.59 \pm 0.08$ & High-res (IGRINS) & \citet{Line2021} \\
      & $0.36^{+0.10}_{-0.09}$ & \textit{JWST} (NIRSpec) & \citet{August2023} \\
      & $0.57\pm0.06$ & \textit{JWST}/NIRSpec + IGRINS combined & \citet{Smith2024} \\
      & $0.54\pm0.12$ & \textit{JWST}/NIRSpec + \textit{HST}/WFC3 combined & \citet{Edwards2024} \\
      \hline
      WASP-127b & $0.56^{+0.05}_{-0.07}$ & High-res (CRIRES+) & \citet{Nortmann2024} \\
      \hline \hline
      Jupiter & $0.90^{+1.54}_{-0.35}$ & C from Galileo, O from Juno & \citet{Wong2004,Li2024} \\
    \end{tabular}
    \caption{List of planets with well-constrained C/O ratios, measured through direct observation of both carbon- and oxygen-bearing species. Note that results which only reported upper or lower limits \citep[e.g.,][]{Brogi2023,Bell2023} are not included here.}
    \label{tab:ctoo}
\end{table*}

\section{Conclusions}
\label{sec:conclude}

We present observations of the ultra-hot Jupiter WASP-76b with Gemini-S/IGRINS between $1.4-2.4$~$\mu$m, reduced with the IGRINS PLP \citep{Sim2014,Lee2016} and analyzed with the newly developed \texttt{IGRINS\_transit} custom pipeline for IGRINS transiting exoplanet data analysis. We detected the presence of H$_{2}$O, CO, and OH in the atmosphere of WASP-76b. 

We performed three sets of retrievals to determine the abundances of these species. First, we retrieved constant-with-altitude abundances for each species and found a volatile metallicity consistent with the stellar metallicity but a significantly super-solar C/O ratio. However, water dissociation in the upper atmospheres of ultra-hot Jupiters is expected to bias retrieved constant-with-altitude abundances \citep[e.g.,][]{Parmentier2018}. We therefore performed a second set of retrievals which included a break pressure parameterization above which abundances dropped to zero to estimate the effect of dissociation in the upper atmosphere. We found that this second retrieval resulted in a stellar metallicity and super-solar C/O ratio
($\left[\frac{\mathrm{C}+\mathrm{O}}{\mathrm{H}}\right]=-0.70^{+1.27}_{-0.93}$ and C/O$=0.80^{+0.07}_{-0.11}$). 
We also retrieved a water break pressure of $\log_{10}P_{b,\mathrm{H}_{2}\mathrm{O}}=-3.28^{+0.94}_{-1.31}$, which as shown in Figure~\ref{fig:equcomp_breakp} is consistent with expectations for the pressure at which the water abundance begins to decrease sharply in a model with the same approximate composition as what we retrieve for WASP-76b. Finally, we performed a gridtrieval using a set of pre-computed self-consistent thermochemical equilibrium models, which slightly favored a sub-stellar metallicity and solar C/O ratio ($\left[\frac{\mathrm{C}+\mathrm{O}}{\mathrm{H}}\right]=-0.74^{+0.23}_{-0.17}$ and C/O$=0.59^{+0.13}_{-0.14}$), but was consistent within $1\sigma$ with the results of the break pressure retrieval. 

Our derived metallicity and C/O ratio are consistent with a wide variety of formation pathways for WASP-76b \citep[e.g.,][]{Madhu2014,Khorshid2022}. When placed in the broader context of all exoplanets with C/O ratios based on simultaneous measurements of carbon- and oxygen-bearing species, the population as a whole seems skewed toward super-solar C/O ratios. Such super-solar C/O ratios could result from accretion beyond the CO snowline \citep[e.g.,][]{Madhu2014,Oberg2016,Madhu2017} or in-situ formation between the soot line and water snow line \citep{Chachan2023}. However, there is significant scatter among the relatively small population of exoplanets with constrained C/O ratios. Trends in the planetary population may become clearer as \textit{JWST} and high-resolution ground-based observations measure precise C/O ratios for a larger sample of planets. Additionally, future observations of WASP-76b with \textit{JWST} would serve to confirm its composition.

In addition to measuring the composition of WASP-76b, we compared the velocities at which H$_{2}$O, CO, and OH were detected to 3D GCM predictions \citep{Wardenier2023}. We found that the velocity offsets observed for CO and OH were relatively consistent with model expectations. However, the H$_{2}$O detection showed a redshifted systemic velocity, which cannot be matched by any of the 3D models of \citet{Wardenier2023}. Future, more precise observations which can resolve the two limbs of the transiting planet separately, or additional models taking into account effects such as magnetic drag \citep{Beltz2023} may help to understand this H$_{2}$O velocity offset.

\begin{acknowledgments}
We thank the anonymous referee for their helpful comments. Based on observations obtained as part of Program IDs GS-2021B-LP-107 and GS-2022B-LP-107 at the international Gemini Observatory, a program of NSF's NOIRLab, which is managed by the Association of Universities for Research in Astronomy (AURA) under a cooperative agreement with the National Science Foundation on behalf of the Gemini Observatory partnership: the National Science Foundation (United States), National Research Council (Canada), Agencia Nacional de Investigaci\'{o}n y Desarrollo (Chile), Ministerio de Ciencia, Tecnolog\'{i}a e Innovaci\'{o}n (Argentina), Minist\'{e}rio da Ci\^{e}ncia, Tecnologia, Inova\c{c}\~{o}es e Comunica\c{c}\~{o}es (Brazil), and Korea Astronomy and Space Science Institute (Republic of Korea). This work used the Immersion Grating Infrared Spectrometer (IGRINS) that was developed under a collaboration between the University of Texas at Austin and the Korea Astronomy and Space Science Institute (KASI) with the financial support of the Mt. Cuba Astronomical Foundation, of the US National Science Foundation under grants AST-1229522 and AST-1702267, of the McDonald Observatory of the University of Texas at Austin, of the Korean GMT Project of KASI, and Gemini Observatory. Support for this work was provided by NASA through the NASA Hubble Fellowship grant HST-HF2-51485.001-A awarded by the Space Telescope Science Institute, which is operated by AURA, Inc., for NASA, under contract NAS5-26555. M.R.L.\ and J.L.B.\ acknowledge support for this work from NSF grant AST-2307177. The authors acknowledge Research Computing at Arizona State University for providing high-performance computing resources that have contributed to the research results reported within this paper.

\end{acknowledgments}

\vspace{5mm}
\facilities{Gemini-S/IGRINS}

\software{IGRINS\_transit \citep{igrins_transit}, astropy \citep{Astropy2013,Astropy2018,Astropy2022}, matplotlib \citep{Hunter2007}, numpy \citep{Harris2020}, pymultinest \citep{Buchner2014}, scipy \citep{Virtanen2021}}

\bibliography{w76}{}

\begin{thebibliography}{}
\expandafter\ifx\csname natexlab\endcsname\relax\def\natexlab#1{#1}\fi
\providecommand{\url}[1]{\href{#1}{#1}}
\providecommand{\dodoi}[1]{doi:~\href{http://doi.org/#1}{\nolinkurl{#1}}}
\providecommand{\doeprint}[1]{\href{http://ascl.net/#1}{\nolinkurl{http://ascl.net/#1}}}
\providecommand{\doarXiv}[1]{\href{https://arxiv.org/abs/#1}{\nolinkurl{https://arxiv.org/abs/#1}}}

\bibitem[{{Astropy Collaboration} {et~al.}(2013){Astropy Collaboration},
  {Robitaille}, {Tollerud}, {Greenfield}, {Droettboom}, {Bray}, {Aldcroft},
  {Davis}, {Ginsburg}, {Price-Whelan}, {Kerzendorf}, {Conley}, {Crighton},
  {Barbary}, {Muna}, {Ferguson}, {Grollier}, {Parikh}, {Nair}, {Unther},
  {Deil}, {Woillez}, {Conseil}, {Kramer}, {Turner}, {Singer}, {Fox}, {Weaver},
  {Zabalza}, {Edwards}, {Azalee Bostroem}, {Burke}, {Casey}, {Crawford},
  {Dencheva}, {Ely}, {Jenness}, {Labrie}, {Lim}, {Pierfederici}, {Pontzen},
  {Ptak}, {Refsdal}, {Servillat}, \& {Streicher}}]{Astropy2013}
{Astropy Collaboration}, {Robitaille}, T.~P., {Tollerud}, E.~J., {et~al.} 2013,
  \aap, 558, A33, \dodoi{10.1051/0004-6361/201322068}

\bibitem[{{Astropy Collaboration} {et~al.}(2018){Astropy Collaboration},
  {Price-Whelan}, {Sip{\H{o}}cz}, {G{\"u}nther}, {Lim}, {Crawford}, {Conseil},
  {Shupe}, {Craig}, {Dencheva}, {Ginsburg}, {VanderPlas}, {Bradley},
  {P{\'e}rez-Su{\'a}rez}, {de Val-Borro}, {Aldcroft}, {Cruz}, {Robitaille},
  {Tollerud}, {Ardelean}, {Babej}, {Bach}, {Bachetti}, {Bakanov}, {Bamford},
  {Barentsen}, {Barmby}, {Baumbach}, {Berry}, {Biscani}, {Boquien}, {Bostroem},
  {Bouma}, {Brammer}, {Bray}, {Breytenbach}, {Buddelmeijer}, {Burke},
  {Calderone}, {Cano Rodr{\'\i}guez}, {Cara}, {Cardoso}, {Cheedella}, {Copin},
  {Corrales}, {Crichton}, {D'Avella}, {Deil}, {Depagne}, {Dietrich}, {Donath},
  {Droettboom}, {Earl}, {Erben}, {Fabbro}, {Ferreira}, {Finethy}, {Fox},
  {Garrison}, {Gibbons}, {Goldstein}, {Gommers}, {Greco}, {Greenfield},
  {Groener}, {Grollier}, {Hagen}, {Hirst}, {Homeier}, {Horton}, {Hosseinzadeh},
  {Hu}, {Hunkeler}, {Ivezi{\'c}}, {Jain}, {Jenness}, {Kanarek}, {Kendrew},
  {Kern}, {Kerzendorf}, {Khvalko}, {King}, {Kirkby}, {Kulkarni}, {Kumar},
  {Lee}, {Lenz}, {Littlefair}, {Ma}, {Macleod}, {Mastropietro}, {McCully},
  {Montagnac}, {Morris}, {Mueller}, {Mumford}, {Muna}, {Murphy}, {Nelson},
  {Nguyen}, {Ninan}, {N{\"o}the}, {Ogaz}, {Oh}, {Parejko}, {Parley}, {Pascual},
  {Patil}, {Patil}, {Plunkett}, {Prochaska}, {Rastogi}, {Reddy Janga},
  {Sabater}, {Sakurikar}, {Seifert}, {Sherbert}, {Sherwood-Taylor}, {Shih},
  {Sick}, {Silbiger}, {Singanamalla}, {Singer}, {Sladen}, {Sooley},
  {Sornarajah}, {Streicher}, {Teuben}, {Thomas}, {Tremblay}, {Turner},
  {Terr{\'o}n}, {van Kerkwijk}, {de la Vega}, {Watkins}, {Weaver}, {Whitmore},
  {Woillez}, {Zabalza}, \& {Astropy Contributors}}]{Astropy2018}
{Astropy Collaboration}, {Price-Whelan}, A.~M., {Sip{\H{o}}cz}, B.~M., {et~al.}
  2018, \aj, 156, 123, \dodoi{10.3847/1538-3881/aabc4f}

\bibitem[{{Astropy Collaboration} {et~al.}(2022){Astropy Collaboration},
  {Price-Whelan}, {Lim}, {Earl}, {Starkman}, {Bradley}, {Shupe}, {Patil},
  {Corrales}, {Brasseur}, {N{\"o}the}, {Donath}, {Tollerud}, {Morris},
  {Ginsburg}, {Vaher}, {Weaver}, {Tocknell}, {Jamieson}, {van Kerkwijk},
  {Robitaille}, {Merry}, {Bachetti}, {G{\"u}nther}, {Aldcroft},
  {Alvarado-Montes}, {Archibald}, {B{\'o}di}, {Bapat}, {Barentsen},
  {Baz{\'a}n}, {Biswas}, {Boquien}, {Burke}, {Cara}, {Cara}, {Conroy},
  {Conseil}, {Craig}, {Cross}, {Cruz}, {D'Eugenio}, {Dencheva}, {Devillepoix},
  {Dietrich}, {Eigenbrot}, {Erben}, {Ferreira}, {Foreman-Mackey}, {Fox},
  {Freij}, {Garg}, {Geda}, {Glattly}, {Gondhalekar}, {Gordon}, {Grant},
  {Greenfield}, {Groener}, {Guest}, {Gurovich}, {Handberg}, {Hart},
  {Hatfield-Dodds}, {Homeier}, {Hosseinzadeh}, {Jenness}, {Jones}, {Joseph},
  {Kalmbach}, {Karamehmetoglu}, {Ka{\l}uszy{\'n}ski}, {Kelley}, {Kern},
  {Kerzendorf}, {Koch}, {Kulumani}, {Lee}, {Ly}, {Ma}, {MacBride}, {Maljaars},
  {Muna}, {Murphy}, {Norman}, {O'Steen}, {Oman}, {Pacifici}, {Pascual},
  {Pascual-Granado}, {Patil}, {Perren}, {Pickering}, {Rastogi}, {Roulston},
  {Ryan}, {Rykoff}, {Sabater}, {Sakurikar}, {Salgado}, {Sanghi}, {Saunders},
  {Savchenko}, {Schwardt}, {Seifert-Eckert}, {Shih}, {Jain}, {Shukla}, {Sick},
  {Simpson}, {Singanamalla}, {Singer}, {Singhal}, {Sinha}, {Sip{\H{o}}cz},
  {Spitler}, {Stansby}, {Streicher}, {{\v{S}}umak}, {Swinbank}, {Taranu},
  {Tewary}, {Tremblay}, {de Val-Borro}, {Van Kooten}, {Vasovi{\'c}}, {Verma},
  {de Miranda Cardoso}, {Williams}, {Wilson}, {Winkel}, {Wood-Vasey}, {Xue},
  {Yoachim}, {Zhang}, {Zonca}, \& {Astropy Project Contributors}}]{Astropy2022}
{Astropy Collaboration}, {Price-Whelan}, A.~M., {Lim}, P.~L., {et~al.} 2022,
  \apj, 935, 167, \dodoi{10.3847/1538-4357/ac7c74}

\bibitem[{{August} {et~al.}(2023){August}, {Bean}, {Zhang}, {Lunine}, {Xue},
  {Line}, \& {Smith}}]{August2023}
{August}, P.~C., {Bean}, J.~L., {Zhang}, M., {et~al.} 2023, \apjl, 953, L24,
  \dodoi{10.3847/2041-8213/ace828}

\bibitem[{{Azevedo Silva} {et~al.}(2022){Azevedo Silva}, {Demangeon}, {Santos},
  {Allart}, {Borsa}, {Cristo}, {Esparza-Borges}, {Seidel}, {Palle}, {Sousa},
  {Tabernero}, {Zapatero Osorio}, {Cristiani}, {Pepe}, {Rebolo}, {Adibekyan},
  {Alibert}, {Barros}, {Bouchy}, {Bourrier}, {Lo Curto}, {Di Marcantonio},
  {D'Odorico}, {Ehrenreich}, {Figueira}, {Gonz{\'a}lez Hern{\'a}ndez}, {Lovis},
  {Martins}, {Mehner}, {Micela}, {Molaro}, {Mounzer}, {Nunes}, {Sozzetti},
  {Su{\'a}rez Mascare{\~n}o}, \& {Udry}}]{Azevedo2022}
{Azevedo Silva}, T., {Demangeon}, O.~D.~S., {Santos}, N.~C., {et~al.} 2022,
  \aap, 666, L10, \dodoi{10.1051/0004-6361/202244489}

\bibitem[{{Bazinet} {et~al.}(2024){Bazinet}, {Pelletier}, {Benneke}, {Salinas},
  \& {Mace}}]{Bazinet2024}
{Bazinet}, L., {Pelletier}, S., {Benneke}, B., {Salinas}, R., \& {Mace}, G.~N.
  2024, \aj, 167, 206, \dodoi{10.3847/1538-3881/ad3071}

\bibitem[{{Bean} {et~al.}(2023){Bean}, {Xue}, {August}, {Lunine}, {Zhang},
  {Thorngren}, {Tsai}, {Stassun}, {Schlawin}, {Ahrer}, {Ih}, \&
  {Mansfield}}]{Bean2023}
{Bean}, J.~L., {Xue}, Q., {August}, P.~C., {et~al.} 2023, \nat, 618, 43,
  \dodoi{10.1038/s41586-023-05984-y}

\bibitem[{{Bedell} {et~al.}(2018){Bedell}, {Bean}, {Mel{\'e}ndez}, {Spina},
  {Ram{\'\i}rez}, {Asplund}, {Alves-Brito}, {dos Santos}, {Dreizler}, {Yong},
  {Monroe}, \& {Casagrande}}]{Bedell2018}
{Bedell}, M., {Bean}, J.~L., {Mel{\'e}ndez}, J., {et~al.} 2018, \apj, 865, 68,
  \dodoi{10.3847/1538-4357/aad908}

\bibitem[{Bell \& Berrington(1987)}]{Bell1987}
Bell, K.~L., \& Berrington, K.~A. 1987, Journal of Physics B: {Atomic \&
  Molecular Physics}, 20, 1

\bibitem[{{Bell} {et~al.}(2023){Bell}, {Welbanks}, {Schlawin}, {Line},
  {Fortney}, {Greene}, {Ohno}, {Parmentier}, {Rauscher}, {Beatty}, {Mukherjee},
  {Wiser}, {Boyer}, {Rieke}, \& {Stansberry}}]{Bell2023}
{Bell}, T.~J., {Welbanks}, L., {Schlawin}, E., {et~al.} 2023, \nat, 623, 709,
  \dodoi{10.1038/s41586-023-06687-0}

\bibitem[{{Beltz} {et~al.}(2023){Beltz}, {Rauscher}, {Kempton}, {Malsky}, \&
  {Savel}}]{Beltz2023}
{Beltz}, H., {Rauscher}, E., {Kempton}, E. M.~R., {Malsky}, I., \& {Savel},
  A.~B. 2023, \aj, 165, 257, \dodoi{10.3847/1538-3881/acd24d}

\bibitem[{{Brogi} \& {Line}(2019)}]{Brogi2019}
{Brogi}, M., \& {Line}, M.~R. 2019, \aj, 157, 114,
  \dodoi{10.3847/1538-3881/aaffd3}

\bibitem[{{Brogi} {et~al.}(2023){Brogi}, {Emeka-Okafor}, {Line}, {Gandhi},
  {Pino}, {Kempton}, {Rauscher}, {Parmentier}, {Bean}, {Mace}, {Cowan},
  {Shkolnik}, {Wardenier}, {Mansfield}, {Welbanks}, {Smith}, {Fortney},
  {Birkby}, {Zalesky}, {Dang}, {Patience}, \& {D{\'e}sert}}]{Brogi2023}
{Brogi}, M., {Emeka-Okafor}, V., {Line}, M.~R., {et~al.} 2023, \aj, 165, 91,
  \dodoi{10.3847/1538-3881/acaf5c}

\bibitem[{{Buchner} {et~al.}(2014{\natexlab{a}}){Buchner}, {Georgakakis},
  {Nandra}, {Hsu}, {Rangel}, {Brightman}, {Merloni}, {Salvato}, {Donley}, \&
  {Kocevski}}]{Buchnerpymulti}
{Buchner}, J., {Georgakakis}, A., {Nandra}, K., {et~al.} 2014{\natexlab{a}},
  \aap, 564, A125, \dodoi{10.1051/0004-6361/201322971}

\bibitem[{{Buchner} {et~al.}(2014{\natexlab{b}}){Buchner}, {Georgakakis},
  {Nandra}, {Hsu}, {Rangel}, {Brightman}, {Merloni}, {Salvato}, {Donley}, \&
  {Kocevski}}]{Buchner2014}
---. 2014{\natexlab{b}}, \aap, 564, A125, \dodoi{10.1051/0004-6361/201322971}

\bibitem[{{Casasayas-Barris} {et~al.}(2021){Casasayas-Barris}, {Orell-Miquel},
  {Stangret}, {Nortmann}, {Yan}, {Oshagh}, {Palle}, {Sanz-Forcada},
  {L{\'o}pez-Puertas}, {Nagel}, {Luque}, {Morello}, {Snellen}, {Zechmeister},
  {Quirrenbach}, {Caballero}, {Ribas}, {Reiners}, {Amado}, {Bergond}, {Czesla},
  {Henning}, {Khalafinejad}, {Molaverdikhani}, {Montes}, {Perger},
  {S{\'a}nchez-L{\'o}pez}, \& {Sedaghati}}]{Casasayas2021}
{Casasayas-Barris}, N., {Orell-Miquel}, J., {Stangret}, M., {et~al.} 2021,
  \aap, 654, A163, \dodoi{10.1051/0004-6361/202141669}

\bibitem[{{Chachan} {et~al.}(2023){Chachan}, {Knutson}, {Lothringer}, \&
  {Blake}}]{Chachan2023}
{Chachan}, Y., {Knutson}, H.~A., {Lothringer}, J., \& {Blake}, G.~A. 2023,
  \apj, 943, 112, \dodoi{10.3847/1538-4357/aca614}

\bibitem[{{de Kok} {et~al.}(2013){de Kok}, {Brogi}, {Snellen}, {Birkby},
  {Albrecht}, \& {de Mooij}}]{deKok2013}
{de Kok}, R.~J., {Brogi}, M., {Snellen}, I.~A.~G., {et~al.} 2013, \aap, 554,
  A82, \dodoi{10.1051/0004-6361/201321381}

\bibitem[{{Deibert} {et~al.}(2021){Deibert}, {de Mooij}, {Jayawardhana},
  {Turner}, {Ridden-Harper}, {Fossati}, {Hood}, {Fortney}, {Flagg},
  {MacDonald}, {Allart}, \& {Sing}}]{Deibert2021}
{Deibert}, E.~K., {de Mooij}, E. J.~W., {Jayawardhana}, R., {et~al.} 2021,
  \apjl, 919, L15, \dodoi{10.3847/2041-8213/ac2513}

\bibitem[{{Deibert} {et~al.}(2023){Deibert}, {de Mooij}, {Jayawardhana},
  {Turner}, {Ridden-Harper}, {Hood}, {Fortney}, {Flagg}, {Fossati}, {Allart},
  {Brogi}, \& {MacDonald}}]{Deibert2023}
---. 2023, \aj, 166, 141, \dodoi{10.3847/1538-3881/acebdc}

\bibitem[{{Edwards} \& {Changeat}(2024)}]{Edwards2024}
{Edwards}, B., \& {Changeat}, Q. 2024, \apjl, 962, L30,
  \dodoi{10.3847/2041-8213/ad2000}

\bibitem[{{Edwards} {et~al.}(2020){Edwards}, {Changeat}, {Baeyens}, {Tsiaras},
  {Al-Refaie}, {Taylor}, {Yip}, {Bieger}, {Blain}, {Gressier}, {Guilluy},
  {Jaziri}, {Kiefer}, {Modirrousta-Galian}, {Morvan}, {Mugnai}, {Pluriel},
  {Poveda}, {Skaf}, {Whiteford}, {Wright}, {Zingales}, {Charnay}, {Drossart},
  {Leconte}, {Venot}, {Waldmann}, \& {Beaulieu}}]{Edwards2020}
{Edwards}, B., {Changeat}, Q., {Baeyens}, R., {et~al.} 2020, \aj, 160, 8,
  \dodoi{10.3847/1538-3881/ab9225}

\bibitem[{{Ehrenreich} {et~al.}(2020){Ehrenreich}, {Lovis}, {Allart}, {Zapatero
  Osorio}, {Pepe}, {Cristiani}, {Rebolo}, {Santos}, {Borsa}, {Demangeon},
  {Dumusque}, {Gonz{\'a}lez Hern{\'a}ndez}, {Casasayas-Barris},
  {S{\'e}gransan}, {Sousa}, {Abreu}, {Adibekyan}, {Affolter}, {Allende Prieto},
  {Alibert}, {Aliverti}, {Alves}, {Amate}, {Avila}, {Baldini}, {Bandy}, {Benz},
  {Bianco}, {Bolmont}, {Bouchy}, {Bourrier}, {Broeg}, {Cabral}, {Calderone},
  {Pall{\'e}}, {Cegla}, {Cirami}, {Coelho}, {Conconi}, {Coretti}, {Cumani},
  {Cupani}, {Dekker}, {Delabre}, {Deiries}, {D'Odorico}, {Di Marcantonio},
  {Figueira}, {Fragoso}, {Genolet}, {Genoni}, {G{\'e}nova Santos}, {Hara},
  {Hughes}, {Iwert}, {Kerber}, {Knudstrup}, {Landoni}, {Lavie}, {Lizon},
  {Lendl}, {Lo Curto}, {Maire}, {Manescau}, {Martins}, {M{\'e}gevand},
  {Mehner}, {Micela}, {Modigliani}, {Molaro}, {Monteiro}, {Monteiro},
  {Moschetti}, {M{\"u}ller}, {Nunes}, {Oggioni}, {Oliveira}, {Pariani},
  {Pasquini}, {Poretti}, {Rasilla}, {Redaelli}, {Riva}, {Santana Tschudi},
  {Santin}, {Santos}, {Segovia Milla}, {Seidel}, {Sosnowska}, {Sozzetti},
  {Span{\`o}}, {Su{\'a}rez Mascare{\~n}o}, {Tabernero}, {Tenegi}, {Udry},
  {Zanutta}, \& {Zerbi}}]{Ehrenreich2020}
{Ehrenreich}, D., {Lovis}, C., {Allart}, R., {et~al.} 2020, \nat, 580, 597,
  \dodoi{10.1038/s41586-020-2107-1}

\bibitem[{{Fortney}(2012)}]{Fortney2012}
{Fortney}, J.~J. 2012, \apjl, 747, L27, \dodoi{10.1088/2041-8205/747/2/L27}

\bibitem[{{Fu} {et~al.}(2021){Fu}, {Deming}, {Lothringer}, {Nikolov}, {Sing},
  {Kempton}, {Ih}, {Evans}, {Stevenson}, {Wakeford}, {Rodriguez}, {Eastman},
  {Stassun}, {Henry}, {L{\'o}pez-Morales}, {Lendl}, {Conti}, {Stockdale},
  {Collins}, {Kielkopf}, {Barstow}, {Sanz-Forcada}, {Ehrenreich}, {Bourrier},
  \& {dos Santos}}]{Fu2021}
{Fu}, G., {Deming}, D., {Lothringer}, J., {et~al.} 2021, \aj, 162, 108,
  \dodoi{10.3847/1538-3881/ac1200}

\bibitem[{{Gagnebin} {et~al.}(2024){Gagnebin}, {Mukherjee}, {Fortney}, \&
  {Batalha}}]{Gagnebin2024}
{Gagnebin}, A., {Mukherjee}, S., {Fortney}, J.~J., \& {Batalha}, N.~E. 2024,
  arXiv e-prints, arXiv:2404.17658, \dodoi{10.48550/arXiv.2404.17658}

\bibitem[{{Gaia Collaboration} {et~al.}(2018){Gaia Collaboration}, {Brown},
  {Vallenari}, {Prusti}, {de Bruijne}, {Babusiaux}, {Bailer-Jones}, {Biermann},
  {Evans}, {Eyer}, {Jansen}, {Jordi}, {Klioner}, {Lammers}, {Lindegren},
  {Luri}, {Mignard}, {Panem}, {Pourbaix}, {Randich}, {Sartoretti}, {Siddiqui},
  {Soubiran}, {van Leeuwen}, {Walton}, {Arenou}, {Bastian}, {Cropper},
  {Drimmel}, {Katz}, {Lattanzi}, {Bakker}, {Cacciari}, {Casta{\~n}eda},
  {Chaoul}, {Cheek}, {De Angeli}, {Fabricius}, {Guerra}, {Holl}, {Masana},
  {Messineo}, {Mowlavi}, {Nienartowicz}, {Panuzzo}, {Portell}, {Riello},
  {Seabroke}, {Tanga}, {Th{\'e}venin}, {Gracia-Abril}, {Comoretto},
  {Garcia-Reinaldos}, {Teyssier}, {Altmann}, {Andrae}, {Audard},
  {Bellas-Velidis}, {Benson}, {Berthier}, {Blomme}, {Burgess}, {Busso},
  {Carry}, {Cellino}, {Clementini}, {Clotet}, {Creevey}, {Davidson}, {De
  Ridder}, {Delchambre}, {Dell'Oro}, {Ducourant},
  {Fern{\'a}ndez-Hern{\'a}ndez}, {Fouesneau}, {Fr{\'e}mat}, {Galluccio},
  {Garc{\'\i}a-Torres}, {Gonz{\'a}lez-N{\'u}{\~n}ez}, {Gonz{\'a}lez-Vidal},
  {Gosset}, {Guy}, {Halbwachs}, {Hambly}, {Harrison}, {Hern{\'a}ndez},
  {Hestroffer}, {Hodgkin}, {Hutton}, {Jasniewicz}, {Jean-Antoine-Piccolo},
  {Jordan}, {Korn}, {Krone-Martins}, {Lanzafame}, {Lebzelter}, {L{\"o}ffler},
  {Manteiga}, {Marrese}, {Mart{\'\i}n-Fleitas}, {Moitinho}, {Mora}, {Muinonen},
  {Osinde}, {Pancino}, {Pauwels}, {Petit}, {Recio-Blanco}, {Richards},
  {Rimoldini}, {Robin}, {Sarro}, {Siopis}, {Smith}, {Sozzetti}, {S{\"u}veges},
  {Torra}, {van Reeven}, {Abbas}, {Abreu Aramburu}, {Accart}, {Aerts},
  {Altavilla}, {{\'A}lvarez}, {Alvarez}, {Alves}, {Anderson}, {Andrei},
  {Anglada Varela}, {Antiche}, {Antoja}, {Arcay}, {Astraatmadja}, {Bach},
  {Baker}, {Balaguer-N{\'u}{\~n}ez}, {Balm}, {Barache}, {Barata}, {Barbato},
  {Barblan}, {Barklem}, {Barrado}, {Barros}, {Barstow}, {Bartholom{\'e}
  Mu{\~n}oz}, {Bassilana}, {Becciani}, {Bellazzini}, {Berihuete}, {Bertone},
  {Bianchi}, {Bienaym{\'e}}, {Blanco-Cuaresma}, {Boch}, {Boeche}, {Bombrun},
  {Borrachero}, {Bossini}, {Bouquillon}, {Bourda}, {Bragaglia}, {Bramante},
  {Breddels}, {Bressan}, {Brouillet}, {Br{\"u}semeister}, {Brugaletta},
  {Bucciarelli}, {Burlacu}, {Busonero}, {Butkevich}, {Buzzi}, {Caffau},
  {Cancelliere}, {Cannizzaro}, {Cantat-Gaudin}, {Carballo}, {Carlucci},
  {Carrasco}, {Casamiquela}, {Castellani}, {Castro-Ginard}, {Charlot},
  {Chemin}, {Chiavassa}, {Cocozza}, {Costigan}, {Cowell}, {Crifo}, {Crosta},
  {Crowley}, {Cuypers}, {Dafonte}, {Damerdji}, {Dapergolas}, {David}, {David},
  {de Laverny}, {De Luise}, {De March}, {de Martino}, {de Souza}, {de Torres},
  {Debosscher}, {del Pozo}, {Delbo}, {Delgado}, {Delgado}, {Di Matteo},
  {Diakite}, {Diener}, {Distefano}, {Dolding}, {Drazinos}, {Dur{\'a}n},
  {Edvardsson}, {Enke}, {Eriksson}, {Esquej}, {Eynard Bontemps}, {Fabre},
  {Fabrizio}, {Faigler}, {Falc{\~a}o}, {Farr{\`a}s Casas}, {Federici},
  {Fedorets}, {Fernique}, {Figueras}, {Filippi}, {Findeisen}, {Fonti},
  {Fraile}, {Fraser}, {Fr{\'e}zouls}, {Gai}, {Galleti}, {Garabato},
  {Garc{\'\i}a-Sedano}, {Garofalo}, {Garralda}, {Gavel}, {Gavras}, {Gerssen},
  {Geyer}, {Giacobbe}, {Gilmore}, {Girona}, {Giuffrida}, {Glass}, {Gomes},
  {Granvik}, {Gueguen}, {Guerrier}, {Guiraud}, {Guti{\'e}rrez-S{\'a}nchez},
  {Haigron}, {Hatzidimitriou}, {Hauser}, {Haywood}, {Heiter}, {Helmi}, {Heu},
  {Hilger}, {Hobbs}, {Hofmann}, {Holland}, {Huckle}, {Hypki}, {Icardi},
  {Jan{\ss}en}, {Jevardat de Fombelle}, {Jonker}, {Juh{\'a}sz}, {Julbe},
  {Karampelas}, {Kewley}, {Klar}, {Kochoska}, {Kohley}, {Kolenberg},
  {Kontizas}, {Kontizas}, {Koposov}, {Kordopatis}, {Kostrzewa-Rutkowska},
  {Koubsky}, {Lambert}, {Lanza}, {Lasne}, {Lavigne}, {Le Fustec}, {Le
  Poncin-Lafitte}, {Lebreton}, {Leccia}, {Leclerc}, {Lecoeur-Taibi},
  {Lenhardt}, {Leroux}, {Liao}, {Licata}, {Lindstr{\o}m}, {Lister}, {Livanou},
  {Lobel}, {L{\'o}pez}, {Managau}, {Mann}, {Mantelet}, {Marchal}, {Marchant},
  {Marconi}, {Marinoni}, {Marschalk{\'o}}, {Marshall}, {Martino}, {Marton},
  {Mary}, {Massari}, {Matijevi{\v{c}}}, {Mazeh}, {McMillan}, {Messina},
  {Michalik}, {Millar}, {Molina}, {Molinaro}, {Moln{\'a}r}, {Montegriffo},
  {Mor}, {Morbidelli}, {Morel}, {Morris}, {Mulone}, {Muraveva}, {Musella},
  {Nelemans}, {Nicastro}, {Noval}, {O'Mullane}, {Ord{\'e}novic},
  {Ord{\'o}{\~n}ez-Blanco}, {Osborne}, {Pagani}, {Pagano}, {Pailler},
  {Palacin}, {Palaversa}, {Panahi}, {Pawlak}, {Piersimoni}, {Pineau}, {Plachy},
  {Plum}, {Poggio}, {Poujoulet}, {Pr{\v{s}}a}, {Pulone}, {Racero}, {Ragaini},
  {Rambaux}, {Ramos-Lerate}, {Regibo}, {Reyl{\'e}}, {Riclet}, {Ripepi}, {Riva},
  {Rivard}, {Rixon}, {Roegiers}, {Roelens}, {Romero-G{\'o}mez}, {Rowell},
  {Royer}, {Ruiz-Dern}, {Sadowski}, {Sagrist{\`a} Sell{\'e}s}, {Sahlmann},
  {Salgado}, {Salguero}, {Sanna}, {Santana-Ros}, {Sarasso}, {Savietto},
  {Schultheis}, {Sciacca}, {Segol}, {Segovia}, {S{\'e}gransan}, {Shih},
  {Siltala}, {Silva}, {Smart}, {Smith}, {Solano}, {Solitro}, {Sordo}, {Soria
  Nieto}, {Souchay}, {Spagna}, {Spoto}, {Stampa}, {Steele},
  {Steidelm{\"u}ller}, {Stephenson}, {Stoev}, {Suess}, {Surdej}, {Szabados},
  {Szegedi-Elek}, {Tapiador}, {Taris}, {Tauran}, {Taylor}, {Teixeira},
  {Terrett}, {Teyssandier}, {Thuillot}, {Titarenko}, {Torra Clotet}, {Turon},
  {Ulla}, {Utrilla}, {Uzzi}, {Vaillant}, {Valentini}, {Valette}, {van Elteren},
  {Van Hemelryck}, {van Leeuwen}, {Vaschetto}, {Vecchiato}, {Veljanoski},
  {Viala}, {Vicente}, {Vogt}, {von Essen}, {Voss}, {Votruba}, {Voutsinas},
  {Walmsley}, {Weiler}, {Wertz}, {Wevers}, {Wyrzykowski}, {Yoldas},
  {{\v{Z}}erjal}, {Ziaeepour}, {Zorec}, {Zschocke}, {Zucker}, {Zurbach}, \&
  {Zwitter}}]{Gaia2018}
{Gaia Collaboration}, {Brown}, A.~G.~A., {Vallenari}, A., {et~al.} 2018, \aap,
  616, A1, \dodoi{10.1051/0004-6361/201833051}

\bibitem[{{Gandhi} {et~al.}(2022){Gandhi}, {Kesseli}, {Snellen}, {Brogi},
  {Wardenier}, {Parmentier}, {Welbanks}, \& {Savel}}]{Gandhi2022}
{Gandhi}, S., {Kesseli}, A., {Snellen}, I., {et~al.} 2022, \mnras, 515, 749,
  \dodoi{10.1093/mnras/stac1744}

\bibitem[{{Gandhi} {et~al.}(2019){Gandhi}, {Madhusudhan}, {Hawker}, \&
  {Piette}}]{Gandhi2019}
{Gandhi}, S., {Madhusudhan}, N., {Hawker}, G., \& {Piette}, A. 2019, \aj, 158,
  228, \dodoi{10.3847/1538-3881/ab4efc}

\bibitem[{{Gandhi} {et~al.}(2023){Gandhi}, {Kesseli}, {Zhang}, {Louca},
  {Snellen}, {Brogi}, {Miguel}, {Casasayas-Barris}, {Pelletier}, {Landman},
  {Maguire}, \& {Gibson}}]{Gandhi2023}
{Gandhi}, S., {Kesseli}, A., {Zhang}, Y., {et~al.} 2023, \aj, 165, 242,
  \dodoi{10.3847/1538-3881/accd65}

\bibitem[{{Gharib-Nezhad} {et~al.}(2021){Gharib-Nezhad}, {Iyer}, {Line},
  {Freedman}, {Marley}, \& {Batalha}}]{Gharib2021}
{Gharib-Nezhad}, E., {Iyer}, A.~R., {Line}, M.~R., {et~al.} 2021, \apjs, 254,
  34, \dodoi{10.3847/1538-4365/abf504}

\bibitem[{{Giacobbe} {et~al.}(2021){Giacobbe}, {Brogi}, {Gandhi}, {Cubillos},
  {Bonomo}, {Sozzetti}, {Fossati}, {Guilluy}, {Carleo}, {Rainer},
  {Harutyunyan}, {Borsa}, {Pino}, {Nascimbeni}, {Benatti}, {Biazzo},
  {Bignamini}, {Chubb}, {Claudi}, {Cosentino}, {Covino}, {Damasso}, {Desidera},
  {Fiorenzano}, {Ghedina}, {Lanza}, {Leto}, {Maggio}, {Malavolta}, {Maldonado},
  {Micela}, {Molinari}, {Pagano}, {Pedani}, {Piotto}, {Poretti}, {Scandariato},
  {Yurchenko}, {Fantinel}, {Galli}, {Lodi}, {Sanna}, \& {Tozzi}}]{Giacobbe2021}
{Giacobbe}, P., {Brogi}, M., {Gandhi}, S., {et~al.} 2021, \nat, 592, 205,
  \dodoi{10.1038/s41586-021-03381-x}

\bibitem[{Gordon {et~al.}(2022)Gordon, Rothman, Hargreaves, Hashemi, Karlovets,
  Skinner, Conway, Hill, Kochanov, Tan, Wcisło, Finenko, Nelson, Bernath,
  Birk, Boudon, Campargue, Chance, Coustenis, Drouin, Flaud, Gamache, Hodges,
  Jacquemart, Mlawer, Nikitin, Perevalov, Rotger, Tennyson, Toon, Tran,
  Tyuterev, Adkins, Baker, Barbe, Canè, Császár, Dudaryonok, Egorov,
  Fleisher, Fleurbaey, Foltynowicz, Furtenbacher, Harrison, Hartmann, Horneman,
  Huang, Karman, Karns, Kassi, Kleiner, Kofman, Kwabia–Tchana, Lavrentieva,
  Lee, Long, Lukashevskaya, Lyulin, Makhnev, Matt, Massie, Melosso,
  Mikhailenko, Mondelain, Müller, Naumenko, Perrin, Polyansky, Raddaoui,
  Raston, Reed, Rey, Richard, Tóbiás, Sadiek, Schwenke, Starikova, Sung,
  Tamassia, Tashkun, {Vander Auwera}, Vasilenko, Vigasin, Villanueva, Vispoel,
  Wagner, Yachmenev, \& Yurchenko}]{Gordon2022}
Gordon, I., Rothman, L., Hargreaves, R., {et~al.} 2022, Journal of Quantitative
  Spectroscopy and Radiative Transfer, 277, 107949,
  \dodoi{https://doi.org/10.1016/j.jqsrt.2021.107949}

\bibitem[{{Guillot}(2010)}]{Guillot2010}
{Guillot}, T. 2010, \aap, 520, A27, \dodoi{10.1051/0004-6361/200913396}

\bibitem[{Harris {et~al.}(2020)Harris, Millman, van~der Walt, Gommers,
  Virtanen, Cournapeau, Wieser, Taylor, Berg, Smith, Kern, Picus, Hoyer, van
  Kerkwijk, Brett, Haldane, del R{\'{i}}o, Wiebe, Peterson,
  G{\'{e}}rard-Marchant, Sheppard, Reddy, Weckesser, Abbasi, Gohlke, \&
  Oliphant}]{Harris2020}
Harris, C.~R., Millman, K.~J., van~der Walt, S.~J., {et~al.} 2020, Nature, 585,
  357, \dodoi{10.1038/s41586-020-2649-2}

\bibitem[{{Hood} {et~al.}(2024){Hood}, {Debras}, {Moutou}, {Klein}, {Tremblin},
  {Parmentier}, {Carmona}, {Meech}, {V{\'e}not}, {Masson}, {Petit}, {Vinatier},
  {Martioli}, {Kiefer}, {Turbet}, \& {ATMOSPHERIX consortium}}]{Hood2024}
{Hood}, T., {Debras}, F., {Moutou}, C., {et~al.} 2024, arXiv e-prints,
  arXiv:2403.19434, \dodoi{10.48550/arXiv.2403.19434}

\bibitem[{Hunter(2007)}]{Hunter2007}
Hunter, J.~D. 2007, Computing in Science \& Engineering, 9, 90,
  \dodoi{10.1109/MCSE.2007.55}

\bibitem[{{John}(1988)}]{John1988}
{John}, T.~L. 1988, A\&A, 193, 189

\bibitem[{{Karman} {et~al.}(2019){Karman}, {Gordon}, {van der Avoird},
  {Baranov}, {Boulet}, {Drouin}, {Groenenboom}, {Gustafsson}, {Hartmann},
  {Kurucz}, {Rothman}, {Sun}, {Sung}, {Thalman}, {Tran}, {Wishnow},
  {Wordsworth}, {Vigasin}, {Volkamer}, \& {van der Zande}}]{Karman2019}
{Karman}, T., {Gordon}, I.~E., {van der Avoird}, A., {et~al.} 2019, \icarus,
  328, 160, \dodoi{10.1016/j.icarus.2019.02.034}

\bibitem[{{Kawauchi} {et~al.}(2022){Kawauchi}, {Narita}, {Sato}, \&
  {Kawashima}}]{Kawauchi2022}
{Kawauchi}, K., {Narita}, N., {Sato}, B., \& {Kawashima}, Y. 2022, \pasj, 74,
  225, \dodoi{10.1093/pasj/psab120}

\bibitem[{{Kesseli} \& {Snellen}(2021)}]{Kesseli2021}
{Kesseli}, A.~Y., \& {Snellen}, I.~A.~G. 2021, \apjl, 908, L17,
  \dodoi{10.3847/2041-8213/abe047}

\bibitem[{{Kesseli} {et~al.}(2022){Kesseli}, {Snellen}, {Casasayas-Barris},
  {Molli{\`e}re}, \& {S{\'a}nchez-L{\'o}pez}}]{Kesseli2022}
{Kesseli}, A.~Y., {Snellen}, I.~A.~G., {Casasayas-Barris}, N., {Molli{\`e}re},
  P., \& {S{\'a}nchez-L{\'o}pez}, A. 2022, \aj, 163, 107,
  \dodoi{10.3847/1538-3881/ac4336}

\bibitem[{{Khorshid} {et~al.}(2022){Khorshid}, {Min}, {D{\'e}sert}, {Woitke},
  \& {Dominik}}]{Khorshid2022}
{Khorshid}, N., {Min}, M., {D{\'e}sert}, J.~M., {Woitke}, P., \& {Dominik}, C.
  2022, \aap, 667, A147, \dodoi{10.1051/0004-6361/202141455}

\bibitem[{{Landman} {et~al.}(2021){Landman}, {S{\'a}nchez-L{\'o}pez},
  {Molli{\`e}re}, {Kesseli}, {Louca}, \& {Snellen}}]{Landman2021}
{Landman}, R., {S{\'a}nchez-L{\'o}pez}, A., {Molli{\`e}re}, P., {et~al.} 2021,
  \aap, 656, A119, \dodoi{10.1051/0004-6361/202141696}

\bibitem[{{Lee} \& {Gullikson}(2016)}]{Lee2016}
{Lee}, J.-J., \& {Gullikson}, K. 2016, {Plp: V2.1 Alpha 3}, v2.1-alpha.3,
  Zenodo,  Zenodo, \dodoi{10.5281/zenodo.56067}

\bibitem[{{Lesjak} {et~al.}(2023){Lesjak}, {Nortmann}, {Yan}, {Cont},
  {Reiners}, {Piskunov}, {Hatzes}, {Boldt-Christmas}, {Czesla}, {Heiter},
  {Kochukhov}, {Lavail}, {Nagel}, {Rains}, {Rengel}, {Rodler}, {Seemann}, \&
  {Shulyak}}]{Lesjak2023}
{Lesjak}, F., {Nortmann}, L., {Yan}, F., {et~al.} 2023, \aap, 678, A23,
  \dodoi{10.1051/0004-6361/202347151}

\bibitem[{{Li} {et~al.}(2024){Li}, {Allison}, {Atreya}, {Brueshaber},
  {Fletcher}, {Guillot}, {Li}, {Lunine}, {Miguel}, {Orton}, {Steffes}, {Waite},
  {Wong}, {Levin}, \& {Bolton}}]{Li2024}
{Li}, C., {Allison}, M., {Atreya}, S., {et~al.} 2024, \icarus, 414, 116028,
  \dodoi{10.1016/j.icarus.2024.116028}

\bibitem[{{Li} {et~al.}(2015){Li}, {Gordon}, {Rothman}, {Tan}, {Hu}, {Kassi},
  {Campargue}, \& {Medvedev}}]{Li2015}
{Li}, G., {Gordon}, I.~E., {Rothman}, L.~S., {et~al.} 2015, \apjs, 216, 15,
  \dodoi{10.1088/0067-0049/216/1/15}

\bibitem[{{Line} {et~al.}(2021){Line}, {Brogi}, {Bean}, {Gandhi}, {Zalesky},
  {Parmentier}, {Smith}, {Mace}, {Mansfield}, {Kempton}, {Fortney}, {Shkolnik},
  {Patience}, {Rauscher}, {D{\'e}sert}, \& {Wardenier}}]{Line2021}
{Line}, M.~R., {Brogi}, M., {Bean}, J.~L., {et~al.} 2021, \nat, 598, 580,
  \dodoi{10.1038/s41586-021-03912-6}

\bibitem[{{Lothringer} {et~al.}(2021){Lothringer}, {Rustamkulov}, {Sing},
  {Gibson}, {Wilson}, \& {Schlaufman}}]{Lothringer2021}
{Lothringer}, J.~D., {Rustamkulov}, Z., {Sing}, D.~K., {et~al.} 2021, \apj,
  914, 12, \dodoi{10.3847/1538-4357/abf8a9}

\bibitem[{{Madhusudhan} {et~al.}(2014){Madhusudhan}, {Amin}, \&
  {Kennedy}}]{Madhu2014}
{Madhusudhan}, N., {Amin}, M.~A., \& {Kennedy}, G.~M. 2014, \apjl, 794, L12,
  \dodoi{10.1088/2041-8205/794/1/L12}

\bibitem[{{Madhusudhan} {et~al.}(2017){Madhusudhan}, {Bitsch}, {Johansen}, \&
  {Eriksson}}]{Madhu2017}
{Madhusudhan}, N., {Bitsch}, B., {Johansen}, A., \& {Eriksson}, L. 2017,
  \mnras, 469, 4102, \dodoi{10.1093/mnras/stx1139}

\bibitem[{{Mansfield} {et~al.}(2021){Mansfield}, {Line}, {Bean}, {Fortney},
  {Parmentier}, {Wiser}, {Kempton}, {Gharib-Nezhad}, {Sing},
  {L{\'o}pez-Morales}, {Baxter}, {D{\'e}sert}, {Swain}, \&
  {Roudier}}]{Mansfield2021}
{Mansfield}, M., {Line}, M.~R., {Bean}, J.~L., {et~al.} 2021, Nature Astronomy,
  5, 1224, \dodoi{10.1038/s41550-021-01455-4}

\bibitem[{{May} {et~al.}(2021){May}, {Komacek}, {Stevenson}, {Kempton}, {Bean},
  {Malik}, {Ih}, {Mansfield}, {Savel}, {Deming}, {Desert}, {Feng}, {Fortney},
  {Kataria}, {Lewis}, {Morley}, {Rauscher}, \& {Showman}}]{May2021}
{May}, E.~M., {Komacek}, T.~D., {Stevenson}, K.~B., {et~al.} 2021, \aj, 162,
  158, \dodoi{10.3847/1538-3881/ac0e30}

\bibitem[{{Molli{\`e}re} {et~al.}(2022){Molli{\`e}re}, {Molyarova}, {Bitsch},
  {Henning}, {Schneider}, {Kreidberg}, {Eistrup}, {Burn}, {Nasedkin},
  {Semenov}, {Mordasini}, {Schlecker}, {Schwarz}, {Lacour}, {Nowak}, \&
  {Schulik}}]{Molliere2022}
{Molli{\`e}re}, P., {Molyarova}, T., {Bitsch}, B., {et~al.} 2022, \apj, 934,
  74, \dodoi{10.3847/1538-4357/ac6a56}

\bibitem[{{Mordasini} {et~al.}(2016){Mordasini}, {van Boekel}, {Molli{\`e}re},
  {Henning}, \& {Benneke}}]{Mordasini2016}
{Mordasini}, C., {van Boekel}, R., {Molli{\`e}re}, P., {Henning}, T., \&
  {Benneke}, B. 2016, \apj, 832, 41, \dodoi{10.3847/0004-637X/832/1/41}

\bibitem[{{Nortmann} {et~al.}(2024){Nortmann}, {Lesjak}, {Yan}, {Cont},
  {Czesla}, {Lavail}, {Rains}, {Nagel}, {Boldt-Christmas}, {Hatzes}, {Reiners},
  {Piskunov}, {Kochukhov}, {Heiter}, {Shulyak}, {Rengel}, \&
  {Seemann}}]{Nortmann2024}
{Nortmann}, L., {Lesjak}, F., {Yan}, F., {et~al.} 2024, arXiv e-prints,
  arXiv:2404.12363, \dodoi{10.48550/arXiv.2404.12363}

\bibitem[{{{\"O}berg} \& {Bergin}(2016)}]{Oberg2016}
{{\"O}berg}, K.~I., \& {Bergin}, E.~A. 2016, \apjl, 831, L19,
  \dodoi{10.3847/2041-8205/831/2/L19}

\bibitem[{{{\"O}berg} {et~al.}(2011){{\"O}berg}, {Murray-Clay}, \&
  {Bergin}}]{Oberg2011}
{{\"O}berg}, K.~I., {Murray-Clay}, R., \& {Bergin}, E.~A. 2011, \apjl, 743,
  L16, \dodoi{10.1088/2041-8205/743/1/L16}

\bibitem[{{Ohno} \& {Fortney}(2023)}]{Ohno2023}
{Ohno}, K., \& {Fortney}, J.~J. 2023, \apj, 946, 18,
  \dodoi{10.3847/1538-4357/acafed}

\bibitem[{{Oliva} {et~al.}(2015){Oliva}, {Origlia}, {Scuderi}, {Benatti},
  {Carleo}, {Lapenna}, {Mucciarelli}, {Baffa}, {Biliotti}, {Carbonaro},
  {Falcini}, {Giani}, {Iuzzolino}, {Massi}, {Sanna}, {Sozzi}, {Tozzi},
  {Ghedina}, {Ghinassi}, {Lodi}, {Harutyunyan}, \& {Pedani}}]{Oliva2015}
{Oliva}, E., {Origlia}, L., {Scuderi}, S., {et~al.} 2015, \aap, 581, A47,
  \dodoi{10.1051/0004-6361/201526291}

\bibitem[{{Parmentier} {et~al.}(2018){Parmentier}, {Line}, {Bean}, {Mansfield},
  {Kreidberg}, {Lupu}, {Visscher}, {D{\'e}sert}, {Fortney}, {Deleuil},
  {Arcangeli}, {Showman}, \& {Marley}}]{Parmentier2018}
{Parmentier}, V., {Line}, M.~R., {Bean}, J.~L., {et~al.} 2018, \aap, 617, A110,
  \dodoi{10.1051/0004-6361/201833059}

\bibitem[{{Pelletier} {et~al.}(2021){Pelletier}, {Benneke}, {Darveau-Bernier},
  {Boucher}, {Cook}, {Piaulet}, {Coulombe}, {Artigau}, {Lafreni{\`e}re},
  {Delisle}, {Allart}, {Doyon}, {Donati}, {Fouqu{\'e}}, {Moutou}, {Cadieux},
  {Delfosse}, {H{\'e}brard}, {Martins}, {Martioli}, \&
  {Vandal}}]{Pelletier2021}
{Pelletier}, S., {Benneke}, B., {Darveau-Bernier}, A., {et~al.} 2021, \aj, 162,
  73, \dodoi{10.3847/1538-3881/ac0428}

\bibitem[{{Pelletier} {et~al.}(2023){Pelletier}, {Benneke}, {Ali-Dib},
  {Prinoth}, {Kasper}, {Seifahrt}, {Bean}, {Debras}, {Klein}, {Bazinet},
  {Hoeijmakers}, {Kesseli}, {Lim}, {Carmona}, {Pino}, {Casasayas-Barris},
  {Hood}, \& {St{\"u}rmer}}]{Pelletier2023}
{Pelletier}, S., {Benneke}, B., {Ali-Dib}, M., {et~al.} 2023, arXiv e-prints,
  arXiv:2306.08739, \dodoi{10.48550/arXiv.2306.08739}

\bibitem[{{Polman} {et~al.}(2023){Polman}, {Waters}, {Min}, {Miguel}, \&
  {Khorshid}}]{Polman2023}
{Polman}, J., {Waters}, L.~B.~F.~M., {Min}, M., {Miguel}, Y., \& {Khorshid}, N.
  2023, \aap, 670, A161, \dodoi{10.1051/0004-6361/202244647}

\bibitem[{{Polyansky} {et~al.}(2018){Polyansky}, {Kyuberis}, {Zobov},
  {Tennyson}, {Yurchenko}, \& {Lodi}}]{Polyansky2018}
{Polyansky}, O.~L., {Kyuberis}, A.~A., {Zobov}, N.~F., {et~al.} 2018, \mnras,
  480, 2597, \dodoi{10.1093/mnras/sty1877}

\bibitem[{{Ramkumar} {et~al.}(2023){Ramkumar}, {Gibson}, {Nugroho}, {Maguire},
  \& {Fortune}}]{Ramkumar2023}
{Ramkumar}, S., {Gibson}, N.~P., {Nugroho}, S.~K., {Maguire}, C., \& {Fortune},
  M. 2023, \mnras, 525, 2985, \dodoi{10.1093/mnras/stad2476}

\bibitem[{{S{\'a}nchez-L{\'o}pez} {et~al.}(2022){S{\'a}nchez-L{\'o}pez},
  {Landman}, {Molli{\`e}re}, {Casasayas-Barris}, {Kesseli}, \&
  {Snellen}}]{SanchezLopez2022}
{S{\'a}nchez-L{\'o}pez}, A., {Landman}, R., {Molli{\`e}re}, P., {et~al.} 2022,
  \aap, 661, A78, \dodoi{10.1051/0004-6361/202142591}

\bibitem[{{Savel} {et~al.}(2023){Savel}, {Kempton}, {Rauscher}, {Komacek},
  {Bean}, {Malik}, \& {Malsky}}]{Savel2023}
{Savel}, A.~B., {Kempton}, E. M.~R., {Rauscher}, E., {et~al.} 2023, \apj, 944,
  99, \dodoi{10.3847/1538-4357/acb141}

\bibitem[{{Savel} {et~al.}(2022){Savel}, {Kempton}, {Malik}, {Komacek}, {Bean},
  {May}, {Stevenson}, {Mansfield}, \& {Rauscher}}]{Savel2022}
{Savel}, A.~B., {Kempton}, E. M.~R., {Malik}, M., {et~al.} 2022, \apj, 926, 85,
  \dodoi{10.3847/1538-4357/ac423f}

\bibitem[{{Schneider} \& {Bitsch}(2021{\natexlab{a}})}]{Schneider2021}
{Schneider}, A.~D., \& {Bitsch}, B. 2021{\natexlab{a}}, \aap, 654, A71,
  \dodoi{10.1051/0004-6361/202039640}

\bibitem[{{Schneider} \& {Bitsch}(2021{\natexlab{b}})}]{Schneider2021b}
---. 2021{\natexlab{b}}, \aap, 654, A72, \dodoi{10.1051/0004-6361/202141096}

\bibitem[{{Seidel} {et~al.}(2019){Seidel}, {Ehrenreich}, {Wyttenbach},
  {Allart}, {Lendl}, {Pino}, {Bourrier}, {Cegla}, {Lovis}, {Barrado},
  {Bayliss}, {Astudillo-Defru}, {Deline}, {Fisher}, {Heng}, {Joseph}, {Lavie},
  {Melo}, {Pepe}, {S{\'e}gransan}, \& {Udry}}]{Seidel2019}
{Seidel}, J.~V., {Ehrenreich}, D., {Wyttenbach}, A., {et~al.} 2019, \aap, 623,
  A166, \dodoi{10.1051/0004-6361/201834776}

\bibitem[{{Sim} {et~al.}(2014){Sim}, {Le}, {Pak}, {Lee}, {Kang}, {Chun},
  {Jeong}, {Yuk}, {Kim}, {Park}, {Pavel}, \& {Jaffe}}]{Sim2014}
{Sim}, C.~K., {Le}, H. A.~N., {Pak}, S., {et~al.} 2014, Advances in Space
  Research, 53, 1647, \dodoi{10.1016/j.asr.2014.02.024}

\bibitem[{{Smith} {et~al.}(2024){Smith}, {Line}, {Bean}, {Brogi}, {August},
  {Welbanks}, {Desert}, {Lunine}, {Sanchez}, {Mansfield}, {Pino}, {Rauscher},
  {Kempton}, {Zalesky}, \& {Fowler}}]{Smith2024}
{Smith}, P. C.~B., {Line}, M.~R., {Bean}, J.~L., {et~al.} 2024, \aj, 167, 110,
  \dodoi{10.3847/1538-3881/ad17bf}

\bibitem[{{Tabernero} {et~al.}(2021){Tabernero}, {Zapatero Osorio}, {Allart},
  {Borsa}, {Casasayas-Barris}, {Demangeon}, {Ehrenreich}, {Lillo-Box}, {Lovis},
  {Pall{\'e}}, {Sousa}, {Rebolo}, {Santos}, {Pepe}, {Cristiani}, {Adibekyan},
  {Allende Prieto}, {Alibert}, {Barros}, {Bouchy}, {Bourrier}, {D'Odorico},
  {Dumusque}, {Faria}, {Figueira}, {G{\'e}nova Santos}, {Gonz{\'a}lez
  Hern{\'a}ndez}, {Hojjatpanah}, {Lo Curto}, {Lavie}, {Martins}, {Martins},
  {Mehner}, {Micela}, {Molaro}, {Nunes}, {Poretti}, {Seidel}, {Sozzetti},
  {Su{\'a}rez Mascare{\~n}o}, {Udry}, {Aliverti}, {Affolter}, {Alves}, {Amate},
  {Avila}, {Bandy}, {Benz}, {Bianco}, {Broeg}, {Cabral}, {Conconi}, {Coelho},
  {Cumani}, {Deiries}, {Dekker}, {Delabre}, {Fragoso}, {Genoni}, {Genolet},
  {Hughes}, {Knudstrup}, {Kerber}, {Landoni}, {Lizon}, {Maire}, {Manescau}, {Di
  Marcantonio}, {M{\'e}gevand}, {Monteiro}, {Monteiro}, {Moschetti}, {Mueller},
  {Modigliani}, {Oggioni}, {Oliveira}, {Pariani}, {Pasquini}, {Rasilla},
  {Redaelli}, {Riva}, {Santana-Tschudi}, {Santin}, {Santos}, {Segovia},
  {Sosnowska}, {Span{\`o}}, {Tenegi}, {Iwert}, {Zanutta}, \&
  {Zerbi}}]{Tabernero2021}
{Tabernero}, H.~M., {Zapatero Osorio}, M.~R., {Allart}, R., {et~al.} 2021,
  \aap, 646, A158, \dodoi{10.1051/0004-6361/202039511}

\bibitem[{{Virtanen} {et~al.}(2021){Virtanen}, {Gommers}, {Burovski},
  {Oliphant}, {Weckesser}, {Cournapeau}, {Alexbrc}, {Reddy}, {Peterson},
  {Haberland}, {Wilson}, {Nelson}, {Endolith}, {Mayorov}, {Van Der Walt},
  {Polat}, {Laxalde}, {Brett}, {Larson}, {Millman}, {Lars}, {Van Mulbregt},
  {Eric-Jones}, {Carey}, {Peterbell10}, {Moore}, {Kern}, {Leslie}, {Perktold},
  \& {Striega}}]{Virtanen2021}
{Virtanen}, P., {Gommers}, R., {Burovski}, E., {et~al.} 2021, {scipy/scipy:
  SciPy 1.6.3}, v1.6.3, Zenodo,  Zenodo, \dodoi{10.5281/zenodo.4718897}

\bibitem[{{von Essen} {et~al.}(2020){von Essen}, {Mallonn}, {Hermansen},
  {Nixon}, {Madhusudhan}, {Kjeldsen}, \&
  {Tautvai{\v{s}}ien{\.{e}}}}]{vonEssen2020}
{von Essen}, C., {Mallonn}, M., {Hermansen}, S., {et~al.} 2020, \aap, 637, A76,
  \dodoi{10.1051/0004-6361/201937169}

\bibitem[{{Wardenier} {et~al.}(2021){Wardenier}, {Parmentier}, {Lee}, {Line},
  \& {Gharib-Nezhad}}]{Wardenier2021}
{Wardenier}, J.~P., {Parmentier}, V., {Lee}, E. K.~H., {Line}, M.~R., \&
  {Gharib-Nezhad}, E. 2021, \mnras, 506, 1258, \dodoi{10.1093/mnras/stab1797}

\bibitem[{{Wardenier} {et~al.}(2023){Wardenier}, {Parmentier}, {Line}, \&
  {Lee}}]{Wardenier2023}
{Wardenier}, J.~P., {Parmentier}, V., {Line}, M.~R., \& {Lee}, E. K.~H. 2023,
  arXiv e-prints, arXiv:2307.04931, \dodoi{10.48550/arXiv.2307.04931}

\bibitem[{{Weiner Mansfield} \& Line(2024)}]{igrins_transit}
{Weiner Mansfield}, M., \& Line, M.~R. 2024, {IGRINS\_transit: analyze
  exoplanet transit observations taken with Gemini-S/IGRINS}, 1.0,  Zenodo,
  \dodoi{10.5281/zenodo.11106414}

\bibitem[{{West} {et~al.}(2016){West}, {Hellier}, {Almenara}, {Anderson},
  {Barros}, {Bouchy}, {Brown}, {Collier Cameron}, {Deleuil}, {Delrez}, {Doyle},
  {Faedi}, {Fumel}, {Gillon}, {G{\'o}mez Maqueo Chew}, {H{\'e}brard}, {Jehin},
  {Lendl}, {Maxted}, {Pepe}, {Pollacco}, {Queloz}, {S{\'e}gransan}, {Smalley},
  {Smith}, {Southworth}, {Triaud}, \& {Udry}}]{West2016}
{West}, R.~G., {Hellier}, C., {Almenara}, J.~M., {et~al.} 2016, \aap, 585,
  A126, \dodoi{10.1051/0004-6361/201527276}

\bibitem[{{Wong} {et~al.}(2004){Wong}, {Mahaffy}, {Atreya}, {Niemann}, \&
  {Owen}}]{Wong2004}
{Wong}, M.~H., {Mahaffy}, P.~R., {Atreya}, S.~K., {Niemann}, H.~B., \& {Owen},
  T.~C. 2004, \icarus, 171, 153, \dodoi{10.1016/j.icarus.2004.04.010}

\bibitem[{{Yan} {et~al.}(2023){Yan}, {Nortmann}, {Reiners}, {Piskunov},
  {Hatzes}, {Seemann}, {Shulyak}, {Lavail}, {Rains}, {Cont}, {Rengel},
  {Lesjak}, {Nagel}, {Kochukhov}, {Czesla}, {Boldt-Christmas}, {Heiter},
  {Smoker}, {Rodler}, {Bristow}, {Dorn}, {Jung}, {Marquart}, \&
  {Stempels}}]{Yan2023}
{Yan}, F., {Nortmann}, L., {Reiners}, A., {et~al.} 2023, \aap, 672, A107,
  \dodoi{10.1051/0004-6361/202245371}

\end{thebibliography}
\bibliographystyle{aasjournal}

\end{document}